\documentclass[journal, a4paper]{IEEEtran}

\usepackage{cite}
\usepackage{amsmath}
\usepackage{algorithmic}
\usepackage{array}
\usepackage{bm}
\usepackage{bbm}
\usepackage{amssymb}
\usepackage{amsthm}
\usepackage{mathrsfs}
\usepackage{empheq}	
\usepackage[mathcal]{euscript}
\usepackage{xcolor}
\usepackage{floatrow}
\usepackage{enumitem}
\usepackage{verbatim}

\newtheoremstyle{bfnote}%
{}{}
{\itshape}{}
{\bfseries}{.}
{ }{\thmname{#1}\thmnumber{ #2}\thmnote{ (#3)}}
\theoremstyle{bfnote}

\newtheorem{theorem}{Theorem}
\newtheorem{lemma}{Lemma}
\newtheorem{corollary}{Corollary}

\newtheorem{assumption}{Assumption}

\newcommand{\E}{\mathbb{E}}
\newcommand{\thetashare}{\theta_{\rm TX}}
\newcommand{\thetatrue}{\theta_0}
\newcommand{\T}{^\top}

\newcommand*{\QED}{\hfill\ensuremath{\square}}%

\newcommand{\indcard}{J}

\usepackage{etoolbox}
\makeatletter
\patchcmd{\@makecaption}
{\scshape}
{}
{}
{}
\makeatother

\usepackage[T1]{fontenc}

\allowdisplaybreaks

\begin{document}
	
	\title{Memory-Aware Social Learning under Partial Information Sharing}
	
	\author{Michele Cirillo, Virginia Bordignon, Vincenzo Matta, and Ali H. Sayed
	\thanks{Michele Cirillo and Vincenzo Matta are with the Department of Information and Electrical Engineering and Applied Mathematics (DIEM), University of Salerno, via Giovanni Paolo II, I-84084, Fisciano (SA), Italy (e-mails: \{micirillo, vmatta\}@unisa.it).
Virginia Bordignon and Ali H. Sayed are with the \'Ecole Polytechnique F\'ed\'erale de Lausanne EPFL, School of Engineering, CH-1015 Lausanne, Switzerland (e-mails: \{virginia.bordignon, ali.sayed\}@epfl.ch).
}
\thanks{
This work was supported in part by SNSF grant 205121-184999. 
A short conference version of this work is under review~\cite{bib:CirilloBordignonMattaSayedICASSP2023}. 
}
	}

	\maketitle
	
\begin{abstract}
This work examines a social learning problem, where dispersed agents connected through a network topology interact locally to form their opinions ({\em beliefs}) as regards certain hypotheses of interest. These opinions evolve over time, since the agents collect observations from the environment, and update their current beliefs by accounting for: their past beliefs, the innovation contained in the new data, and the beliefs received from the neighbors.
The distinguishing feature of the present work is that agents are constrained to share opinions regarding only a single hypothesis. 
We devise a novel learning strategy where each agent forms a valid belief by completing the partial beliefs received from its neighbors. This completion is performed by exploiting the knowledge accumulated in the past beliefs, thanks to a principled {\em memory-aware} rule inspired by a Bayesian criterion.
The analysis allows us to characterize the role of memory in social learning under {\em partial information sharing}, revealing novel and nontrivial learning dynamics. 
Surprisingly, we establish that the standard classification rule based on selecting the maximum belief is not optimal under partial information sharing, while there exists a consistent threshold-based decision rule that allows each agent to classify correctly the hypothesis of interest. 
We also show that the proposed strategy outperforms previously considered schemes, highlighting that the introduction of memory in the social learning algorithm is critical to overcome the limitations arising from sharing partial information. 
%
\end{abstract} 

	
\section{Introduction}
Inspired by realistic social dynamics, social learning refers to a family of algorithmic strategies through which agents update and propagate {\em beliefs} within a network~\cite{acemoglu2011opinion,acemoglu2011bayesian,jadbabaie2012non,zhao2012learning,chamley2013models,krishnamurthy2013social,mossel2017opinion,bib:nedic2017,molavi2018theory,salami2017social,bib:lalitha2018,bib:matta2020,bordignon2021adaptive}. All agents observe a common phenomenon of interest, which can be explained by discrete-valued states or {\em hypotheses}. The belief of an agent summarizes its opinion regarding these hypotheses. In order to update its belief, each agent collects evidence from the phenomenon of interest in the form of streaming observations, which depend on the unknown true state of nature. Conceived initially as a model for opinion formation within groups, social learning has evolved to encompass engineered decision-making systems. An example of a system is a network of connected weather sensors measuring different meteorological attributes and updating their confidence regarding possible weather forecasts, e.g., ``imminent rain'', ``clear skies'', ``thunderstorm''.
		
{\em Non-Bayesian} social learning solutions~\cite{jadbabaie2012non,zhao2012learning,bib:nedic2017,molavi2018theory,salami2017social,bib:lalitha2018,bib:matta2020,bordignon2021adaptive} overcome the exceeding computational complexity associated with fully Bayesian approaches~\cite{mossel2013making,hkazla2021bayesian} and allow agents to aggregate information in a two-step procedure: {\em i)} a local Bayesian update is used to incorporate information contained in the stream of observations; and {\em ii)} a combination step incorporates the knowledge contained in the beliefs of neighboring agents.
		In all these strategies, a common assumption is that each agent has access to the {\em full} beliefs of all neighbors. However, this condition is not verified in several contexts.

		From a behavioral point of view, it is often the case that individuals limit their shared opinions to a {\em single} candidate state. For example, consider the process of choosing the best product among brands $\{\theta_1,\theta_2,\theta_3\}$. 
		A new product has been recently released by brand $\theta_1$, and then a group of individuals interacts by exchanging reviews regarding the new release. The information contained in these reviews is limited to a single candidate product and ignores the remaining hypotheses $\theta_2,\theta_3$. 
		
		Partial sharing of information is also motivated from an engineering point of view, as we are increasingly interested in designing communication-efficient systems~\cite{bib:koloskova2019, bib:mills2020}. 
		Moreover, withholding information from other agents is often motivated by privacy reasons and regulation constraints.
		
		In the realm of communication-efficient social learning, we can distinguish distinct simultaneous efforts. In~\cite{bib:toghani2022}, the authors propose sharing quantized belief ratios to decrease the communication load. In~\cite{bib:mitra2020}, the authors propose an event-triggered algorithm, which reduces communication to instants when there is sufficient innovation in the beliefs. Other approaches used to reduce communication in social learning allow agents to exchange beliefs with only one randomly-sampled component or neighbor at a time~\cite{bib:kayaalpbordignonsayed2022, bib:inan2022}.
		
		A {\em partial information} strategy for social learning is introduced in~\cite{bib:bordignon2020socialicassp, bib:bordignon2020social}, where it is assumed that agents can only transmit their beliefs concerning a {\em single} hypothesis of interest denoted by $\thetashare$. In the proposed solution, upon receiving partial beliefs, agents have to fill in the missing information regarding the non-transmitted components, and they do so by adopting a uniform approach, i.e., by assuming the non-transmitted belief components have equal mass. In the present work, we propose alternatively that the local {\em memory} of each agent should be used to fill in the missing belief entries of its neighbors. 
		
In this work, we characterize the learning behavior of the agents under the proposed strategy. We show that when the shared hypothesis is {\em not} the true one, all agents are able to correctly discard it with full confidence, i.e., the belief regarding $\thetashare$ converges to $0$ as the number of observations grows. 
		The key condition required in this case is the classical {\em global identifiability} assumption, which prescribes that the network as a whole possesses sufficient information to solve the inference problem, even if the agents are unable to learn well individually.~\cite{bib:bordignon2020socialicassp,bib:bordignon2020social}. 
		When the shared hypothesis is equal to the true one, we first show that agents achieve full confidence around the truth, i.e., the belief regarding $\thetashare$ converges to $1$, if there exists {\em at least one} agent that would be able to learn in isolation. 
		Then, we establish that the existence of such a powerful agent is actually unnecessary since, under the simpler condition of global identifiability, we show that there exists a decision rule that allows us to classify correctly the nature (true/false) of the hypothesis of interest.
		Remarkably, this rule is not based on maximization of the belief, as is common practice in the literature. 
		We actually ascertain that under partial information sharing, standard maximization of the belief is {\em not} optimal for achieving correct classification.
		We further show that the proposed strategy outperforms the existing schemes. 

		\textbf{Notation.} We use boldface fonts to represent random variables. The symbols $\xrightarrow{\rm a.s.}$ and $ \xrightarrow{\rm p}$ indicate respectively almost sure convergence and convergence in probability~\cite{bib:billingsley}, as the time index $i$ goes to infinity.

\section{Partial Information Sharing}
Consider a network of $K$ agents, which are trying to identify the true state of nature $\thetatrue$ from a set of discrete hypotheses $\Theta\triangleq \{1,2,\dots, H\}$. To solve the inference problem, each agent $k$ receives at instant $i$ an observation $\boldsymbol{\xi}_{k,i}\in\mathcal{X}_k$ and possesses a set of likelihoods $L_k(\xi|\theta)$ for $\xi\in\mathcal{X}_k$ and $\theta\in\Theta$, which are agent-dependent models for the distribution of observations $\boldsymbol{\xi}_{k,i}$ given each hypothesis $\theta$. 
Technically, $L_k(\xi | \theta)$ is a likelihood function when regarded as a function of $\theta$. It is instead a probability density or mass function (depending on whether the observations are continuous or discrete, respectively) when regarded as a function of $\xi$. The observations $\boldsymbol{\xi}_{k,i}$ are distributed according to the true model $L_k(\xi|\thetatrue)$ and are assumed to be independent and identically distributed (iid) over time, i.e., over $i$, but can be dependent across agents, i.e., over $k$.

		The network is represented by a graph, in which nodes play
		the role of agents and the edges symbolize the communication
		links between agents. Each of these edges, e.g., from agent $\ell$
		to agent $k$, is assigned a nonnegative weight $a_{\ell k}$, which
		quantifies the confidence that agent $k$ has in the information
		coming from its neighbor $\ell$. These weights can be conveniently arranged into a left-stochastic matrix $A=[a_{\ell k}]$ satisfying:
		\begin{equation}
			\mathbbm{1}\T A= \mathbbm{1}\T,\quad  a_{\ell k}\geq 0, \quad a_{\ell k}=0 \text{ if } \ell\notin \mathcal{N}_k,
			\label{eq:A}
		\end{equation}
		where $\mathcal{N}_k$ denotes the set of neighbors of agent $k$, including agent $k$ itself.
		
		Traditional social learning allows the network of agents to update and propagate opinions (or beliefs) regarding the possible hypotheses and to learn the truth, i.e., to find the true state $\thetatrue$, in a decentralized manner~\cite{jadbabaie2012non,zhao2012learning,bib:nedic2017,molavi2018theory,bib:lalitha2018}. 
		At each instant $i$, agent $k$ updates its belief vector
		$\boldsymbol{\mu}_{k,i}$, which is a probability mass function over the set of possible hypotheses, i.e., $\boldsymbol{\mu}_{k,i}\in\Delta^H$, where $\Delta^H$ denotes the probability simplex with dimension $H$. Its $\theta$-th component $\boldsymbol{\mu}_{k,i}(\theta)$ quantifies how
		confident agent $k$ is at instant $i$ that $\theta$ is the true state of
		nature.

		In traditional social learning, each agent $k$ starts with its own initial belief vector $\mu_{k,0}$ and, for $i=1,2,\dots$, iteratively updates its belief vector $\bm{\mu}_{k,i}$ using the following two-step procedure:
\begin{subequations}
		\begin{align}
			\bm{\psi}_{k,i}(\theta)&\propto L_k(\bm{\xi}_{k,i}|\theta)\bm{\mu}_{k,i-1}(\theta),
			\label{eq:model-psi-first}
			\\
			\bm{\mu}_{k,i}(\theta)&\propto
			\prod\limits_{\ell\in\mathcal{N}_k}[\bm{\psi}_{\ell,i}(\theta)]^{a_{\ell k}}.
			\label{eq:model-mu-old}
		\end{align}
		\end{subequations}
		As is standard in the Bayesian framework, the proportionality sign hides the proportionality constant that is necessary to guarantee the conditions:
		$\sum_{\theta\in\Theta}\bm{\psi}_{k,i}(\theta)=1$ and $\sum_{\theta\in\Theta}\bm{\mu}_{k,i}(\theta)=1$.
Step~\eqref{eq:model-psi-first} is a local Bayesian update. Specifically, agent $k$ implements Bayes' rule to build an {\em intermediate belief vector} $\bm{\psi}_{k,i}$, by combining: {\em i)} the prior information contained in the previous belief vector $\bm{\mu}_{k,i-1}$; and {\em ii)} the likelihood $L_k(\bm{\xi}_{k,i}|\theta)$ that exploits the fresh information contained in $\bm{\xi}_{k,i}$. The intermediate beliefs are subsequently transmitted across the network. In step~\eqref{eq:model-mu-old}, agent $k$ updates the belief vector by combining its own intermediate belief vector $\bm{\psi}_{k,i}$ and the ones received from its neighborhood $\mathcal{N}_k$ into a weighted geometric average with combination weights $a_{\ell k}$. 
		
		The algorithm described above has been thoroughly studied in the literature~\cite{bib:nedic2017,bib:lalitha2018,bib:matta2020}. It has been shown
		that, under some mild technical conditions, all agents learn the truth with probability one as $i\rightarrow\infty$, that is, $\boldsymbol{\mu}_{k,i}(\thetatrue)\xrightarrow{\rm a.s.} 1$.

One limitation of traditional social learning is that at each iteration $i$ each agent $k$ receives from its neighbors the {\em complete} belief vectors $\{\bm{\psi}_{\ell,i}\}_{\ell\in\mathcal{N}_k}$.
Motivated, e.g., by behavioral, communication, or privacy constraints, agents might not be willing to share their entire belief vectors. To address this scenario, a variation of this algorithm has been proposed in~\cite{bib:bordignon2020social}. In this new setting, agents share a {\em single} component of their beliefs, corresponding to some {\em hypothesis of interest} $\thetashare\in\Theta$.
	Formally, the new algorithm consists of the following three steps, iterated by each agent $k$ for $i=1,2,\ldots$:
\begin{subequations}
	\begin{align}
		\bm{\psi}_{k,i}(\theta)&\propto L_k(\bm{\xi}_{k,i}|\theta)\bm{\mu}_{k,i-1}(\theta)
		\label{eq:model-psi}\\
		\widehat{\bm{\psi}}_{\ell k,i}&={\sf F}_k \big(\bm{\psi}_{\ell,i}(\thetashare)\big), \textnormal{ for $\ell\in\mathcal{N}_k$}
		\label{eq:sfT}\\
		\bm{\mu}_{k, i} (\theta) &\propto \prod_{\ell \in\mathcal{N}_k}  [\widehat{\bm{\psi}}_{\ell k, i} (\theta)]^{a_{\ell k}}
		\label{eq:model-mu}
	\end{align}
	\end{subequations}
Step \eqref{eq:model-psi} is still a Bayesian update. The main novelty is in step \eqref{eq:sfT}: agent $\ell\in\mathcal{N}_k$ sends the belief component $\bm{\psi}_{\ell, i}(\thetashare)$ to agent $k$, which, in turn, builds a complete {\em estimated} belief vector $\widehat{\bm{\psi}}_{\ell k,i}$. 
	This is the estimated belief vector constructed by agent $k$ {\em relative} to its neighbor $\ell$, in the sense that, starting from the received component $\bm{\psi}_{\ell,i}(\thetashare)$, agent $k$ fills the missing entries according to a local transformation ${\sf F}_k:\mathbb{R} \mapsto \Delta^H$. 
	We remark that in our treatment ${\sf F}_k$ will be allowed to be a {\em random} transformation as it can depend on the random beliefs possessed by agent $k$.
	Finally, step \eqref{eq:model-mu} performs the same averaging operation as \eqref{eq:model-mu-old}, but using the estimated beliefs $\widehat{\bm{\psi}}_{\ell k,i}$ in place of the complete beliefs $\bm{\psi}_{k,i}$.
	
\begin{figure}[t]
		\centering
		\includegraphics[width=3.5in]{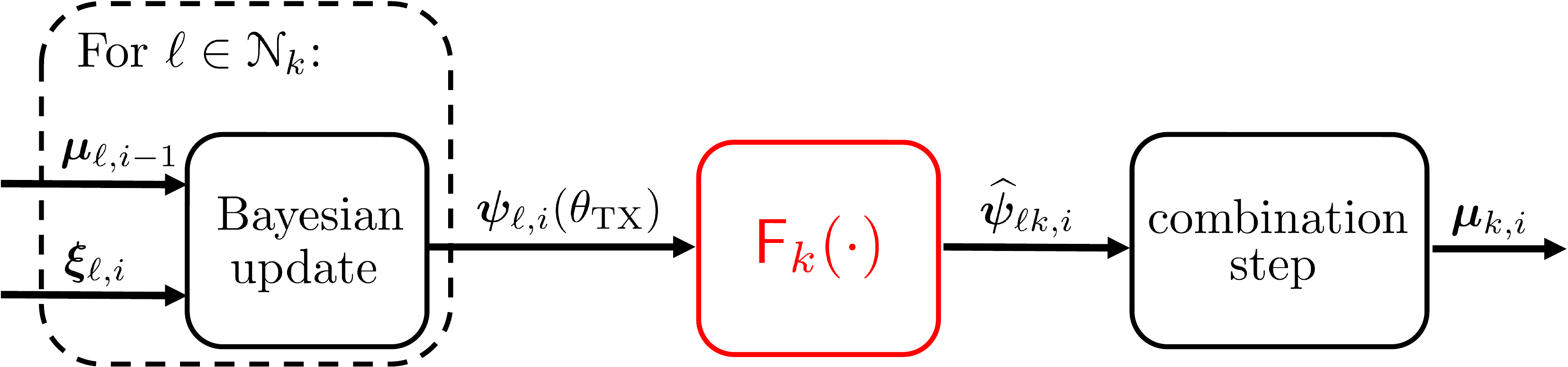}
		\caption{Diagram of the social learning algorithm with partial information. }
		\label{fig:partialdiag}
	\end{figure}

	Figure~\ref{fig:partialdiag} shows a block diagram of this procedure, highlighting in red the novelties introduced by the partial information scenario. The choice of the filling strategy ${\sf F}_k$ is critical to achieve correct learning.
	
\subsection{Filling Strategies}
We now show how the filling strategy can be derived from a Bayesian approach. 
Agent $k$ receives from agent $\ell\in\mathcal{N}_k$ the intermediate belief component $\bm{\psi}_{\ell,i}(\thetashare)$. In the construction of the estimated belief $\widehat{\bm{\psi}}_{\ell k,i}$, agent $k$ trusts agent $\ell$ and, hence, it sets $\widehat{\bm{\psi}}_{\ell k,i}(\thetashare)=\bm{\psi}_{\ell,i}(\thetashare)$.
Once we assume this equality, the remaining mass assigned to the set $\mathcal{T}\triangleq\{\theta\neq\thetashare\}$ must necessarily be $1-\bm{\psi}_{\ell,i}(\thetashare)$. 
From Bayes’ rule, this implies that the belief $\widehat{\bm{\psi}}_{\ell k,i}$ must fulfill, for all $\theta\neq\thetashare$, the equation:\footnote{To interpret \eqref{eq:Bayesfilling}, consider a random variable $\bm{\theta}\in\Theta$. For all $\theta\in\mathcal{T}$:
\begin{equation}
\mathbb{P}[\bm{\theta}=\theta]=\mathbb{P}[\bm{\theta}=\theta, \bm{\theta}\in\mathcal{T}]=
\mathbb{P}[\bm{\theta}=\theta | \bm{\theta}\in\mathcal{T}] \, \mathbb{P}[\bm{\theta}\in\mathcal{T}], 
\label{eq:Bayesxplain}
\end{equation}
where the first equality holds since $\theta\in\mathcal{T}$, while the second equality is Bayes' rule. 
We see from \eqref{eq:Bayesxplain} that the belief of a particular value $\theta\in\mathcal{T}$ can be expressed as the product of a conditional belief times the total belief assigned to set $\mathcal{T}$.
}
\begin{equation}
\widehat{\bm{\psi}}_{\ell k,i}(\theta)=\bm{b}(\theta|\mathcal{T})(1-\bm{\psi}_{\ell,i}(\thetashare)),
\label{eq:Bayesfilling}
\end{equation}
where $\bm{b}(\theta|\mathcal{T})$ is the belief {\em conditioned} on set $\mathcal{T}$.
To complete the filling strategy, it is necessary to choose the form of $\bm{b}(\theta|\mathcal{T})$.

In~\cite{bib:bordignon2020social}, the agnostic maximum-entropy choice
\begin{equation}
\bm{b}(\theta|\mathcal{T})=\frac{1}{H-1},
\label{eq:agnostic}
\end{equation}
is proposed, where agent $k$ assumes no knowledge available to determine $\bm{b}(\theta|\mathcal{T})$. 
In this work, we propose instead to exploit the most up-to-date knowledge that agent $k$ has accumulated up to time $i$. As a matter of fact, the most up-to-date belief available at agent $k$ at time $i$ is $\bm{\psi}_{k,i}$, which leads to the {\em conditional} belief given $\mathcal{T}$:
\begin{equation}
\bm{b}(\theta|\mathcal{T})=\frac{\bm{\psi}_{k,i}(\theta)}{1-\bm{\psi}_{k,i}(\thetashare)}.
\label{eq:memoryconditional}
\end{equation}
We see that \eqref{eq:memoryconditional} diversifies the allocation of the conditional belief mass across the non-transmitted hypotheses, based on the available knowledge stored in the intermediate belief vector $\bm{\psi}_{k,i}$. 
In contrast, strategy \eqref{eq:agnostic} opts for a uniform allocation, thus forgetting any evidence that agent $k$ accumulated in the past. 
We shall accordingly refer to \eqref{eq:agnostic} as the {\em memoryless} strategy, and to \eqref{eq:memoryconditional} as the {\em memory-aware} strategy.

Note also that with strategy \eqref{eq:memoryconditional} agent $k$ is automatically {\em self-aware}, which means that for $\ell=k$ we get $\widehat{\bm{\psi}}_{kk,i}=\bm{\psi}_{k,i}$.
Conservation of the belief available at agent $k$ is a compelling property, and it is interesting to remark that it arises naturally from our Bayesian interpretation of the filling strategy, once we allow it to incorporate the information contained in $\bm{\psi}_{k,i}$. 
In comparison, note that in strategy \eqref{eq:agnostic} agent $k$ is {\em not} self-aware.

In preparation for the forthcoming analysis, it is convenient to write compactly the two obtained strategies:
\\

\noindent
\underline{{\bf Memoryless filling strategy}}
\begin{equation}
		\widehat{\bm{\psi}}_{\ell k, i} (\theta)=
		\begin{cases}
			\bm{\psi}_{\ell, i} (\thetashare), \qquad
			&\theta = \thetashare,
			\\
			\displaystyle{\frac{1}{H-1}} \Big(1 -\bm{\psi}_{\ell, i}(\thetashare)\Big), \qquad
			& \theta \neq \thetashare.
		\label{eq:partialagnost}
\end{cases}
\end{equation}
\underline{{\bf Memory-aware filling strategy}}
\begin{equation}
		\widehat{\bm{\psi}}_{\ell k, i} (\theta)=
		\begin{cases}
			\bm{\psi}_{\ell, i} (\thetashare), \qquad
			&\theta = \thetashare,
			\\
			\displaystyle{\frac{\bm{\psi}_{ k, i} (\theta)}{1 -\bm{\psi}_{k, i}(\thetashare)}}
			\Big(1 -\bm{\psi}_{\ell, i}(\thetashare)\Big), \qquad
			& \theta \neq \thetashare.
		\label{eq:model-memorypsi2}
\end{cases}
\end{equation}

\section{Assumptions}
In this section we collect the assumptions involved in our analysis. We start with a standard assumption on the network connectivity.	
	\begin{assumption}[Strongly-Connected Network] \label{th:strong}
The network is strongly connected, which means that, given any pair of nodes $(\ell,k)$, paths with nonzero weights exist in both directions, i.e., from $\ell$ to $k$ and vice versa (the two paths need not be the same) and that at least one agent $k$ in the entire network has a positive self-weight ($a_{kk}>0$).
		\QED
	\end{assumption}
	Since $A$ is left-stochastic, strong connectivity implies that $A$ is a primitive matrix~\cite{bib:matrix,sayedbook}. Then, the Perron-Frobenius theorem implies the existence of a vector $v$, a.k.a. the Perron eigenvector, which satisfies the following conditions~\cite{bib:matrix,sayedbook}:
	\begin{equation}
		Av=v,\quad \mathbbm{1}\T v=1, \quad v\succ 0,
	\end{equation}
	where $\succ$ indicates element-wise strict inequality.

Next we introduce two regularity conditions involving the likelihood models and the initial beliefs.
	\begin{assumption}[Finiteness of KL Divergences] \label{th:finiteKL}
		For each $k=1,2,\dots, K$ and each pair of hypotheses $\theta,\theta^\prime$,
\begin{equation}
			D_{\rm KL}( L_{k,\theta}||L_{k,\theta^\prime})<\infty.
		\end{equation}
		\QED
	\end{assumption}
	 
	\begin{assumption}[Positive Initial Beliefs]
		\label{th:positivemu0}
		For each $k=1,2,\dots, K$ and each $\theta\in\Theta$, $\mu_{k,0}(\theta)>0$.
		\QED
	\end{assumption}
	We remark that Assumption~\ref{th:finiteKL} implies that the likelihood $L_k(\bm{\xi}_{k,i}|\theta)$ can be equal to zero only for an ensemble of realizations $\bm{\xi}_{k,i}$ having zero probability. Accordingly, starting from Assumption~\ref{th:positivemu0}, we see inductively that, with probability $1$, the beliefs $\bm{\psi}_{k,i}(\theta)$ and $\bm{\mu}_{k,i}(\theta)$ are strictly positive:  {\em i)} from~\eqref{eq:model-psi} we see that, if $\bm{\mu}_{k,i-1}(\theta) > 0$, then $\bm{\psi}_{k,i}(\theta)>0$ with probability $1$; and {\em ii)} from~\eqref{eq:model-mu}, since the combination weights are nonnegative and convex, we have $\bm{\mu}_{k,i}(\theta)>0$ with probability $1$.

We continue by introducing the classical assumption of {\em global identifiability} adopted in social learning.
Let us define, for each agent $k$, the {\em set of indistinguishable hypotheses}:
	\begin{equation}
		\mathcal{D}_k\triangleq \left\{\theta\in\Theta\, : \, D_{\rm KL}( L_{k,\thetatrue}||L_{k,\theta})\neq 0\right\},
	\end{equation}
	where $D_{\rm KL}(\cdot\|\cdot)$ denotes the Kullback-Leibler (KL) divergence~\cite{cover1999elements}. 
	Note that, for ease of notation, when working with KL divergences, we use the notation $L_{k,\theta}$ in place of $L_k(\xi|\theta)$.	
Likewise, the {\em set of indistinguishable hypotheses} is:
	\begin{equation}
		\mathcal{I}_k\triangleq \left\{\theta\in\Theta\setminus\{\thetatrue\}\, : \, D_{\rm KL}( L_{k,\thetatrue}||L_{k,\theta})=0\right\}.
	\end{equation}
In practice, the fact that an agent has two identical likelihoods $L_{k,\thetatrue}$ and $L_{k,\theta}$ means that its data $\{\bm{\xi}_{k,i}\}$ originate from the same mechanism under $\thetatrue$ or $\theta$. For example, if a sensor can only measure the amplitude of a sinusoidal signal, its measurements will have the same characteristics under two distinct signal phases $\thetatrue$ and $\theta$.
We note that agents with nonempty indistinguishable sets are unable to solve the inference problem on their own. 
One advantage of social learning algorithms is that agents can overcome their local identifiability limitations by solving the inference problem in a collaborative manner. In particular, identifiability limitations of the {\em individual} agents can be overcome at the {\em network} level: if agent $k$ is able to distinguish one hypothesis $\theta$ from the true one (i.e., if $\theta\in\mathcal{D}_k$), then the entire network can potentially take advantage of this ability, including some other agent $\ell$ for which $\theta\in\mathcal{I}_\ell$.
These concepts are summarized in the assumption of global identifiability.
\begin{assumption}[Global Identifiability] \label{th:gi}
For each $\theta \neq \thetatrue$, there exists at least one agent $k$ for which $\theta \in \mathcal{D}_k$, which means that $\bigcap_{k=1}^K \mathcal{I}_k=\emptyset$.
\QED
\end{assumption}

	Finally, the next assumption rules out the case where at some agent the likelihood of the true hypothesis is equal to a convex combination of the likelihoods of the distinguishable hypotheses. 
\begin{assumption}[Likelihood Models] \label{th:convexcomb}
For all agents whose set of distinguishable hypotheses $\mathcal{D}_k$ is nonempty, the likelihood $L_{k,\theta_0}$ is not a convex combination of the likelihoods of the distinguishable hypotheses, i.e., for all vectors $\alpha=[\alpha(\theta)]_{\theta\in\mathcal{D}_k}$ with nonnegative entries that add up to $1$, we have:
\begin{equation}
L_{k,\theta_0}\neq
\sum\limits_{\theta\in\mathcal{D}_{k}}\alpha(\theta)L_{k,\theta}.
\label{eq:DKLposalpha}
\end{equation}
\QED
\end{assumption}
Assumption~\ref{th:convexcomb} is a sufficient condition that is useful to prove our results. 
It is typically verified when the agents employ parametric families of likelihoods, where different hypotheses are identified by different values of the parameters. For example, in a Gaussian, exponential, or binomial family it is not possible to represent one likelihood within the class as the convex combination of other likelihoods in the same class. 
To gain further insight on how likely it is to violate Assumption~\ref{th:convexcomb}, it is also useful to consider the following {\em unstructured} setting, where the likelihoods are chosen in a completely random manner. 
Specifically, let $\mathcal{X}_k$ be a discrete finite space, and recall that a probability mass function on $\mathcal{X}_k$ is a point lying in the probability simplex $\Delta_{|\mathcal{X}_k|}$. Assume that the likelihoods are picked uniformly at random from the probability simplex. 
We now show that when the cardinality of the observation space $\mathcal{X}_k$ is larger than the cardinality of the distinguishable set $\mathcal{D}_k$, the probability of picking a set of likelihoods that violate Assumption~\ref{th:convexcomb} is zero. 
This is because the dimensionality of the probability simplex is $d_1=|\mathcal{X}_k|-1$, whereas the dimensionality of the convex hull generated by $|\mathcal{D}_k|$ likelihoods is at most $d_2=|\mathcal{D}_k|-1<d_1$. Therefore, if we pick some points uniformly at random in a continuous space of dimensionality $d_1$, the probability that they fall into a space of dimensionality $d_2<d_1$ is zero.
%

\section{Main Theorems}
	In this section we study the asymptotic behavior of algorithm~\eqref{eq:model-psi}--\eqref{eq:model-mu} when adopting the memory-aware strategy in~\eqref{eq:model-memorypsi2}. 

	In the following treatment we will use the following notation:
	\begin{equation}
		\bm{\mu}_{k,i}(\mathcal{S}) \triangleq \sum_{\theta\in \mathcal{S}} \bm{\mu}_{k,i}(\theta),\qquad 
		\bm{\psi}_{k,i}(\mathcal{S}) \triangleq \sum_{\theta\in \mathcal{S}} \bm{\psi}_{k,i}(\theta),
		\label{eq:subsetS}
	\end{equation}
	for any subset $\mathcal{S}\subseteq \Theta$, with the convention that $\bm{\mu}_{k,i}(\emptyset)=\bm{\psi}_{k,i}(\emptyset)=0$.
	We are now ready to present our main results.

	Preliminarily, it is useful to introduce some descriptors of the learning problem that will be useful to state our results. 
We define for each agent $k$ the ratio:
		\begin{equation}
			\rho_k \triangleq \frac{\mu_{k,0}(\mathcal{I}_k)}{\mu_{k,0}(\thetatrue)}.
			\label{eq:confuratio}
		\end{equation}
The ratio $\rho_k$ quantifies the initial displacement between the mass assigned by agent $k$ to the indistinguishable hypotheses w.r.t. to the true hypothesis. 
High values of $\rho_k$ are thus not beneficial for the agent, since they tend to favor the indistinguishable hypotheses. Conversely, small values of $\rho_k$ favor the true hypothesis $\theta_0$. Exploiting \eqref{eq:subsetS}, we see that the uniform prior assignment $\mu_{k,0}(\theta)=1/H$ for all $\theta\in\Theta$ leads to:
		\begin{equation}
					\rho_k = \frac{\mu_{k,0}(\mathcal{I}_k)}{\mu_{k,0}(\thetatrue)}=
					\frac{\sum_{\theta\in\mathcal{I}_k}(1/H)}{(1/H)}=
					\indcard_k,
		\end{equation} 
		namely, with a flat prior the ratio $\rho_k$ coincides with the cardinality of the indistinguishable set $\indcard_k\triangleq|\mathcal{I}_k|$. 
		According to the above interpretation, we will refer to $\rho_k$ as the {\em confusion ratio}.
		
		In our {\em social} learning environment, it is also useful to consider a measure of confusion at the {\em network} level.
		We accordingly introduce the {\em network confusion ratio}:
		\begin{equation}
		\rho\triangleq \prod_{k=1}^K \rho^{v_k}_k,
		\label{eq:confunet}
		\end{equation}
		which is a weighted geometric average of the individual confusion ratios $\{\rho_k\}$, with weights given by the entries of the Perron eigenvector. Likewise, we introduce the weighted geometric average of the cardinalities of the indistinguishable sets:
		\begin{equation}
		\indcard\triangleq \prod_{k=1}^K \indcard^{v_k}_k.
		\label{eq:confunetcard}
		\end{equation}
Before stating our main theorems, it is useful to recall that under partial information sharing the agents exchange information relative to a single hypothesis of interest $\thetashare$. Accordingly, under partial information sharing we will say that a social learning strategy learns well if each agent is able to establish whether $\thetashare$ is true or false~\cite{bib:bordignon2020social}.

Our first theorem examines the learning behavior of the memory-aware strategy when $\thetashare\neq\thetatrue$.

	\begin{theorem}[Belief Convergence when $\thetashare\neq\thetatrue$] \label{th:main-TXneq0}
		Let $\thetashare\neq\theta_{0}$ and let Assumptions~\ref{th:strong}--\ref{th:convexcomb} hold. 
		Then, for all $k=1,2,\ldots,K$ we have the following properties:
		\begin{itemize}
			\item {\bf Transmitted hypothesis}.
			\begin{equation}
				\bm{\mu}_{k, i} (\thetashare) \xrightarrow{\rm a.s.} 0.
				\label{eq:TXlim}
			\end{equation}
			\item 
			{\bf Non-transmitted, distinguishable hypotheses $\theta\in\mathcal{D}_k\setminus\{\thetashare\}$}.
			\begin{equation}
				\bm{\mu}_{k, i} (\theta) \xrightarrow{\rm a.s.} 0.
				\label{eq:0-again}
			\end{equation}
			\item 
			{\bf True hypothesis and non-transmitted, indistinguishable hypotheses $\theta \in \{\theta_{0}\} \cup \big(\mathcal{I}_k \setminus\{\thetashare\} \big)$}.

			\vspace*{5pt}
			Letting $\mathcal{I}_{k,\overline{\rm TX}}\triangleq\{\theta_{0}\} \cup \big(\mathcal{I}_k \setminus\{\thetashare\} \big)$, from \eqref{eq:TXlim} and \eqref{eq:0-again} we have that the total belief of $\mathcal{I}_{k,\overline{\rm TX}}$ accumulates all the residual mass:
			\begin{equation}
				\bm{\mu}_{k, i} (\mathcal{I}_{k,\overline{\rm TX}}) \xrightarrow{\rm a.s.} 1.
				\label{eq:residualtotal}
			\end{equation}
			Moreover, the allocation of this mass preserves the initial conditional belief of $\theta$ given $\mathcal{I}_{k,\overline{\rm TX}}$:
			\begin{equation}
				\frac{\bm{\mu}_{k, i} (\theta)}{\bm{\mu}_{k, i} (\mathcal{I}_{k,\overline{\rm TX}})} = \frac{\mu_{k, 0} (\theta)}{\mu_{k, 0} (\mathcal{I}_{k,\overline{\rm TX}})}.
				\label{eq:residual}
			\end{equation}

		\end{itemize}
	\end{theorem}
	\begin{IEEEproof} See Appendix~\ref{sec:main-TXneq0-proof}.
	\end{IEEEproof}


The fundamental message from Theorem~\ref{th:main-TXneq0} is that all agents are able to learn well when $\thetashare\neq\thetatrue$, since they end up placing zero mass on the (false) transmitted hypothesis.
In addition, the theorem shows that all distinguishable non-transmitted hypotheses are discarded, and that the whole mass is then distributed over the true and the indistinguishable non-transmitted hypotheses.
Equation \eqref{eq:residual} reveals that the conditional beliefs given these hypotheses are {\em static}, i.e., they are equal to the conditional {\em prior} beliefs. 
This makes perfect sense, since: $i)$ the observations cannot help agent $k$ distinguish between the true and the indistinguishable hypotheses; and $ii)$ no information on the non-transmitted  hypotheses is diffused across the network. 
Therefore, regarding the true and the indistinguishable non-transmitted hypotheses, the information available to agent $k$ is in fact the same information present {\em at the beginning of the learning process}.


Let us switch to the case $\thetashare=\thetatrue$, which is covered by the next theorem.

	\begin{theorem}[Belief Convergence when $\thetashare=\thetatrue$] \label{th:main-TXeq0}
		Let $\thetashare=\theta_{0}$, let $\rho$ be the network confusion ratio defined in \eqref{eq:confunet}, and let Assumptions~\ref{th:strong}--\ref{th:convexcomb} hold. 
		Then, for all $k=1,2,\ldots,K$ we have the following properties:
		\begin{itemize}
			\item {\bf Transmitted hypothesis}.
			\begin{equation}
				\bm{\mu}_{k, i} (\thetashare) \xrightarrow{\rm a.s.} \frac{1}{1+\rho}.
				\label{eq:const-lim}
			\end{equation}
			\item 
			{\bf Distinguishable hypotheses $\theta\in\mathcal{D}_k$}.
			\begin{equation}
				\bm{\mu}_{k, i} (\theta) \xrightarrow{\rm a.s.} 0.
				\label{eq:0}
			\end{equation}
			\item 
			{\bf Indistinguishable hypotheses $\theta \in \mathcal{I}_k$}.
	
			\vspace*{5pt}		
			From \eqref{eq:const-lim} and \eqref{eq:0}, the total belief of $\mathcal{I}_{k}$ accumulates the residual mass:
			\begin{equation}
				\bm{\mu}_{k, i} (\mathcal{I}_k) \xrightarrow{\rm a.s.} \frac{\rho}{1+\rho}.
			\label{eq:residualindistinguish}
			\end{equation}
			Moreover, the allocation of this mass preserves the {\em initial conditional} belief of $\theta$ given $\mathcal{I}_{k}$:
			\begin{equation}
				\frac{\bm{\mu}_{k, i} (\theta)}{\bm{\mu}_{k, i} (\mathcal{I}_k)}
				 = \frac{\mu_{k, 0} (\theta)}{\mu_{k, 0} (\mathcal{I}_k)}.
				\label{eq:1-c}
			\end{equation}
		\end{itemize}
	\end{theorem}

	\begin{IEEEproof}
		See Appendix~\ref{sec:main-TXeq0-proof}.
	\end{IEEEproof}

Theorem~\ref{th:main-TXeq0} provides insightful formulas to capture the learning mechanism of the memory-aware strategy. We now examine the main conclusions revealed by these formulas.  
In traditional social learning with full information sharing, the belief of the true hypothesis converges to $1$ as $i\rightarrow\infty$.
In order to reproduce the same behavior under partial information sharing, we see from \eqref{eq:const-lim} that the network confusion ratio $\rho$ must be zero.
From \eqref{eq:confunet} we know that $\rho$ is a weighted geometric average of the single-agent confusion ratios $\{\rho_k\}$ defined in \eqref{eq:confuratio}. 
Accordingly, $\rho$ can be zero, provided that at least one agent $k$ has $\rho_k=0$, which means that the indistinguishable set $\mathcal{I}_k$ is empty. 
This does not mean that the problem must be locally identifiable at {\em any} agent; this must be the case for just {\em one} powerful, clear-sighted agent. 
For example, we can have a problem that is {\em not} locally identifiable at $K-1$ agents, which are unable to discriminate $\thetatrue$ from some of the other hypotheses. By exploiting cooperation across the network, these agents can profit from the strong agent and overcome their individual limitations.

When such a powerful agent does not exist, we have instead $\rho>0$. 
In this case, while \eqref{eq:0} reveals that zero mass is still assigned to the distinguishable hypotheses, the residual mass is now split between $\theta_0$ {\em and} the indistinguishable hypotheses, since the belief of the true hypothesis converges to a value strictly less than $1$. 
This splitting is ruled by \eqref{eq:0} and \eqref{eq:1-c}, implying that the behavior of the memory-aware social learning strategy {\em depends on the initial beliefs} and in particular that different initial beliefs at different agents can lead to different behavior across agents. 
This conclusion is in contrast with traditional social learning, whose asymptotic behavior is instead independent of the initial belief (provided that nonzero mass is assigned to all hypotheses). 
However, in the last discussion we compared social learning under partial information sharing against traditional social learning by implicitly focusing on the requirement that the belief of $\thetashare$ converges to $1$ when $\thetashare=\thetatrue$. Is this condition really necessary to classify correctly $\thetashare$? We will answer this fundamental question in Sec.~\ref{sec:decisionrule}. Before doing that, it is useful to gain further insights by examining the interesting case of unbiased initialization.

\subsection{Unbiased Initialization}
The most interesting scenario to capture the authentic learning mechanism of the memory-aware strategy is the {\em unbiased} case where the initial beliefs are all uniform. It is useful to summarize the results for this case in the following corollary.

	\begin{corollary}[Belief Convergence with Uniform Initial Beliefs] \label{th:main-TXeq0-coroll}
	Let Assumptions~\ref{th:strong}--\ref{th:convexcomb} hold, and consider, for each agent $k$, the uniform prior assignment $\mu_{k,0}(\theta)=1/H$ for all $\theta\in\Theta$. 
	
	When $\thetashare\neq\theta_{0}$, Eqs. \eqref{eq:TXlim} and \eqref{eq:0-again} hold as they are. Equations \eqref{eq:residualtotal} and \eqref{eq:residual} specialize to, for all $k=1,2,\ldots,K$:
			\begin{equation}
				\bm{\mu}_{k, i} (\theta) \xrightarrow{\rm a.s.} \frac{1}{|\mathcal{I}_{k,\overline{\rm TX}}|},
				\label{eq:residualuniform}
			\end{equation}				
			i.e., the mass is asymptotically equipartitioned over the set comprising $\thetatrue$ and the non-transmitted indistinguishable hypotheses.
		
		When $\thetashare=\theta_{0}$, for all $k=1,2,\ldots,K$ we have the following properties.
		\begin{itemize}
			\item {\bf True hypothesis}.
			\begin{equation}
				\bm{\mu}_{k, i} (\thetatrue) \xrightarrow{\rm a.s.} \frac{1}{1+\indcard}.
				\label{eq:const-lim-coroll}
			\end{equation}
			\item {\bf Distinguishable hypotheses $\theta\in\mathcal{D}_k$}.
			\begin{equation}
				\bm{\mu}_{k, i} (\theta) \xrightarrow{\rm a.s.} 0.
				\label{eq:0-coroll}
			\end{equation}
			\item {\bf Indistinguishable hypotheses $\theta \in \mathcal{I}_k$}.
			\begin{equation}
				\bm{\mu}_{k, i} (\theta)
				 \xrightarrow{\rm a.s.} \left(\frac{\indcard}{\indcard_k}\right)\,\frac{1}{1+\indcard}.
				\label{eq:1-c-coroll}
			\end{equation}
		\end{itemize}
	\end{corollary}
	\begin{IEEEproof}
		The claims follow from Theorems~\ref{th:main-TXneq0} and~\ref{th:main-TXeq0} by setting $\mu_{k,0}(\theta)=1/H$ for all $\theta\in\Theta$ and all $k=1,2,\ldots,K$.	
	\end{IEEEproof}
Corollary~\ref{th:main-TXeq0-coroll} allows us to investigate more closely the role of cooperation in the memory-aware strategy. 
The particular case $\indcard = 0$ (i.e., $\rho=0$) has been discussed in the comments on Theorem~\ref{th:main-TXneq0}.  
Let us then focus on the case $\indcard>0$.

Observe that the geometric average of a set of numbers is bounded by the minimum and maximum values in the set. 
Recalling that $\indcard$ is an average cardinality, in the network we have in general agents with $\indcard_k>\indcard$ and agents with $\indcard_k\leq\indcard$. 
Consider first the agents with a number of indistinguishable hypotheses $\indcard_k$ larger than the average $\indcard$, i.e., with $(\indcard/\indcard_k)<1$.
In view of \eqref{eq:const-lim-coroll} and \eqref{eq:1-c-coroll}, this implies that, with probability $1$ for sufficiently large $i$:
\begin{equation}
\indcard_k>\indcard:\quad \bm{\mu}_{k, i} (\thetatrue)>\bm{\mu}_{k, i} (\theta),~ \textnormal{for all $\theta\in\Theta\setminus\{\theta_0\}$}.
\label{eq:moreconfused}
\end{equation}
Conversely, agents with $(\indcard/\indcard_k)\geq 1$ end up with:
\begin{equation}
\indcard_k\leq \indcard:\quad \bm{\mu}_{k, i} (\thetatrue)\leq \bm{\mu}_{k, i} (\theta),~ \textnormal{for all $\theta\in\Theta\setminus\{\theta_0\}$}.
\label{eq:lessconfused}
\end{equation}
This behavior has an interesting implication on the role of cooperation. 
We see that, after cooperation, the agents that were {\em individually more confused} ($\indcard_k>J$, see \eqref{eq:moreconfused}) truly benefit from cooperation, and end up with a belief that is maximized at the true hypothesis. 
However, the situation is reversed for the agents that were {\em individually less confused} ($\indcard_k< J$, see \eqref{eq:lessconfused}) but then end up with a belief that is no longer maximized at the true hypothesis. 
Note that in the boundary case $\indcard_k=\indcard$ we have multiple indistinguishable maxima.
In summary, cooperation appears to be not beneficial for all agents, as the belief is not necessarily maximized at the true hypothesis. 
However, the aforementioned discussion is based on the implicit assumption that maximization of the beliefs is what one should aim for in order to classify the transmitted hypothesis. 
Is maximization of the belief really necessary to classify correctly $\thetashare$? The next section will provide an unexpected answer to this question.


\section{A Fundamental Dichotomy in Memory-Aware Social Learning}
\label{sec:decisionrule}
Examining jointly Theorems~\ref{th:main-TXneq0} and~\ref{th:main-TXeq0}, we can explain the learning behavior of the memory-aware social learning strategy and the reason why a dichotomy arises between the cases $\thetashare\neq\thetatrue$ and $\thetashare=\thetatrue$.

Consider first the case $\thetashare\neq\thetatrue$.
For what concerns the distinguishable non-transmitted hypotheses $\theta\in\mathcal{D}_k\setminus\{\thetashare\}$, agent $k$ would be able to distinguish them even if working in isolation, by definition of $\mathcal{D}_k$. 
We conclude from \eqref{eq:0-again} that the social learning strategy does not affect this ability of agent $k$, which remains able to discard any $\theta\in\mathcal{D}_k\setminus\{\thetashare\}$ in the distributed context.
For what concerns the transmitted hypothesis $\thetashare\neq\thetatrue$, agent $k$ working in isolation would not be able to discard it properly if it were indistinguishable. On the other hand, we know from global identifiability that there exists at least one agent in the network that is able to distinguish $\thetashare$ from the true hypothesis and, hence, to discard it. Equation \eqref{eq:TXlim} reveals that, thanks to information diffusion across the network, the ability of this particular agent is transferred to {\em all agents}, which become able to discard the transmitted hypothesis with full confidence.

Consider now the case $\thetashare=\thetatrue$.
The behavior regarding the distinguishable hypotheses is the same as before --- see \eqref{eq:0}. 
Also the behavior of the indistinguishable hypotheses is the same as before --- see \eqref{eq:1-c}.
Regarding the true transmitted hypothesis $\thetashare=\thetatrue$, we see from \eqref{eq:const-lim} that the belief of $\thetashare$ converges to a positive quantity, albeit not necessarily to $1$, nor to the maximum belief.

Collecting the results for $\thetashare\neq\thetatrue$ and $\thetashare=\thetatrue$, we see the following fundamental dichotomy arising:
\begin{corollary}[Asymptotic Classification of $\thetashare$] 
\label{th:TXornotcoroll}
Under Assumptions~\ref{th:strong}--\ref{th:convexcomb}, the memory-aware strategy satisfies the following convergences:
\begin{equation}
\begin{cases}
\bm{\mu}_{k,i}(\thetashare)
\xrightarrow{\rm a.s.} 0, \qquad &\textnormal{if $\thetashare\neq\thetatrue$},
\\
\\
\bm{\mu}_{k,i}(\thetashare)
\xrightarrow{\rm a.s.} \displaystyle{\frac{1}{1+\rho}},&\textnormal{if $\thetashare=\thetatrue$}.
\end{cases}
\label{eq:fundichot}
\end{equation}
\end{corollary}
\begin{IEEEproof}
The claim follows from \eqref{eq:TXlim} and \eqref{eq:const-lim}.
\end{IEEEproof}

The asymptotic strictly-positive gap exhibited by the belief of $\thetashare$ under the two possible cases suggests that it is always possible to devise a decision rule that makes each agent $k$ capable of classifying correctly $\thetashare$ with probability $1$ as $i\rightarrow\infty$.
More precisely, we need to define a decision rule for each time $i$, and examine the online behavior of the resulting decisions as $i\rightarrow\infty$.
When the belief of $\thetashare$ converges to $1$ if $\thetashare=\thetatrue$ and to $0$ otherwise, correct classification of the transmitted hypothesis is obvious. 
For example, the standard rule that selects the hypothesis maximizing the belief ends up correctly accepting $\thetashare$ when it is true and rejecting it otherwise.
However, we have observed that if $\bm{\mu}_{k,i}(\thetashare)$ does not converge to $1$ when $\thetashare=\thetatrue$, the rule maximizing the belief can fail in this case, since the maximum mass can be allocated on the indistinguishable hypotheses. 

We propose instead to employ the following threshold test:
\begin{equation}
\begin{cases}
\bm{\mu}_{k,i}(\thetashare)\leq\tau \Rightarrow \textnormal{reject $\thetashare$},\\
\bm{\mu}_{k,i}(\thetashare)>\tau \Rightarrow \textnormal{accept $\thetashare$},
\end{cases}
\qquad 0<\tau<\frac{1}{1+\rho},
\label{eq:decrule}
\end{equation}
which, in view of \eqref{eq:fundichot}, guarantees that the probability of classifying correctly $\thetashare$ converges to $1$ as $i\rightarrow\infty$. 

For rule \eqref{eq:decrule} to work properly, it is necessary to ensure that $\tau$ is smaller than $1/(1+\rho)$. 
From definition \eqref{eq:confunet}, we see that to compute $\rho$ each agent needs to know the initial belief assignments of all agents, as well as the Perron eigenvector. When this knowledge is available, the threshold can be surely set. However, there are several situations where this knowledge is not available. 
We now show that it is possible to set a threshold $\tau<1/(1+\rho)$ with a much coarser prior information. 
To this end, we start by observing from \eqref{eq:confuratio} that we can write:
\begin{equation}
\rho_k=\frac{\mu_{k,0}(\mathcal{I}_k)}{\mu_{k,0}(\thetatrue)}
\leq
\frac{1-\mu_{k,0}(\thetatrue)}{\mu_{k,0}(\thetatrue)}
=
\frac{1}{\mu_{k,0}(\thetatrue)}-1
\leq
\frac{1}{\mu_{\min,0}}-1,
\label{eq:rhokbound}
\end{equation}
where 
\begin{equation}
\mu_{\min,0}=\min_{\substack{k=1,2,\ldots,K\\ \theta\in\Theta}}\mu_{k,0}(\theta).
\end{equation}
The average confusion ratio $\rho$ is upper bounded by the maximum confusion ratio across the agents, which in view of \eqref{eq:rhokbound} yields:
\begin{equation}
\rho\leq \frac{1}{\mu_{\min,0}} - 1\Rightarrow \mu_{\min,0}\leq \frac{1}{1+\rho},
\label{eq:rhonetbound}
\end{equation}
further implying that the choice
\begin{equation}
\tau=\mu_{\min,0}-\varepsilon, \qquad \textnormal{with $\varepsilon<\mu_{\min,0}$},
\label{eq:tausetting}
\end{equation} 
guarantees that $\tau<1/(1+\rho)$.
Accordingly, with \eqref{eq:tausetting} the transmitted hypothesis is accepted provided that the observed belief exceeds (but for a small $\varepsilon$) the smallest initial belief of all agents and all hypotheses. 
In particular, in the case of unbiased initialization Eq. \eqref{eq:tausetting} becomes:
\begin{equation}
\tau=\frac 1 H-\varepsilon, \qquad \textnormal{with $\varepsilon<\frac 1 H$},
\label{eq:fracHminusepsi}
\end{equation}  
which essentially means that a belief larger than the uniform belief is sufficient to accept the transmitted hypothesis. 

Note that to implement \eqref{eq:tausetting}, the agents must know $\mu_{\min,0}$. 
This requires, for example, that the agents share their initial beliefs in a preliminary phase of the algorithm, or that the initial beliefs are assigned with a protocol known to all agents beforehand. 
Remarkably, with an unbiased initialization, see \eqref{eq:fracHminusepsi}, the only quantity necessary to set the threshold is the number of hypotheses, which is obviously known to all agents.

Before concluding this section, it is useful to contrast the truth-learning concept employed in traditional social learning with the decision rule \eqref{eq:decrule}.  
In both cases, each agent is able to make the right choice with probability $1$ as $i\rightarrow\infty$. 
However, there is a difference that can emerge depending on the particular application context.
Following traditional social learning, we might require that the belief $\bm{\mu}_{k,i}(\thetashare)$ converges to $1$ or $0$ if the transmitted hypothesis is true or false, respectively. 
This viewpoint is important in applications where the agents are humans or animals, since it reflects the natural behavior by which individuals express the strength of their opinions, and this strength is expected to increase as more evidence is collected. In particular, the choice of accepting the transmitted hypothesis can be naturally formulated in terms of selecting the maximum belief.

On the other side, when allowing $\rho>0$ in Theorem~\ref{th:main-TXeq0}, the situation changes, since the belief of the transmitted hypothesis is allowed to be even smaller than the belief of some indistinguishable hypothesis. 
This notwithstanding, we showed that the decision rule \eqref{eq:decrule} allows to achieve correct decisions. 
This is because to accept $\thetashare$ this rule does not require that $\bm{\mu}_{k,i}(\thetashare)$ converges to $1$, nor that it is the maximum belief! We showed that it is sufficient to fulfill the milder requirement that $\bm{\mu}_{k,i}(\thetashare)$ exceeds the minimum initial belief.
One explanation for this behavior is as follows. 
When $\thetashare=\thetatrue$, the shared hypothesis is by definition not statistically different from the {\em indistinguishable} hypotheses. 
However, what makes the shared hypothesis different from the indistinguishable hypotheses is the way it is treated  by the social learning algorithm, since it is the only hypothesis the agents exchange information about. This induces the agents to treat $\thetashare$ in a ``privileged'' way. In other words, by using the decision rule \eqref{eq:decrule} in place of the maximum-belief rule, the agents introduce a bias in favor of $\thetashare$, which is used to overcome the limitations of partial information sharing. 
This requires that the agents are aware of how the underlying algorithm works, since to learn correctly they must combine this additional knowledge with their beliefs. 
From a practical viewpoint, this is definitely possible when agents are programmable devices.

\section{Comparison Against Other Schemes}
		\subsection{Comparison Against the Standalone Algorithm}
		Another fundamental aspect is how the social collaboration influences the choices of the agents. To this end, we shall compare the proposed learning algorithm with the standalone algorithm, where there is no information exchange and the agents iteratively update their beliefs as:
\begin{equation}
			\bm{\mu}_{k,i}(\theta)\propto L_k(\bm{\xi}_{k,i}|\theta)\bm{\mu}_{k,i-1}(\theta).
			\label{eq:nosocial}
		\end{equation}
The results achieved by the standalone scheme can be derived using similar analytical tools as used in~\cite{bib:degroot,bib:matta2020}, and we report them without proof in the following theorem.

	\begin{theorem}[Standalone Learning Algorithm] \label{th:standalone}
				Consider the standalone learning algorithm~\eqref{eq:nosocial}, and let $\rho_k$ be the confusion ratio of agent $k$ defined in \eqref{eq:confuratio}.
				Then, under Assumptions~\ref{th:finiteKL} and~\ref{th:positivemu0}, we have the following properties holding for all $k=1,2,\ldots,K$:
		\begin{itemize}
			\item {\bf True hypothesis}.
			\begin{equation}
				\bm{\mu}_{k, i} (\thetatrue) \xrightarrow{\rm a.s.} \frac{1}{1+\rho_k}.
				\label{eq:const-lim-standalone}
			\end{equation}
			\item 
			{\bf Distinguishable hypotheses $\theta\in\mathcal{D}_k$}.
			\begin{equation}
				\bm{\mu}_{k, i} (\theta) \xrightarrow{\rm a.s.} 0.
				\label{eq:0-standalone}
			\end{equation}
			\item {\bf Indistinguishable hypotheses $\theta \in \mathcal{I}_k$}.
			From \eqref{eq:const-lim} and \eqref{eq:0} we have that the total belief of $\mathcal{I}_{k}$ accumulates the residual mass:
			\begin{equation}
				\bm{\mu}_{k, i} (\mathcal{I}_k) \xrightarrow{\rm a.s.} \frac{\rho_k}{1+\rho_k}.
			\label{eq:residualindistinguish-standalone}
			\end{equation}
			Moreover, the allocation of this mass preserves the {\em initial conditional} belief of $\theta$ given $\mathcal{I}_{k}$:
			\begin{equation}
				\frac{\bm{\mu}_{k, i} (\theta)}{\bm{\mu}_{k, i} (\mathcal{I}_k)}
				 =\frac{\mu_{k, 0} (\theta)}{\mu_{k, 0} (\mathcal{I}_k)}.
				\label{eq:1-c-standalone}
			\end{equation}
			\end{itemize}
	\end{theorem}
\hfill\ensuremath{\blacksquare}

We see a nice symmetry between Theorems~\ref{th:main-TXeq0} and~\ref{th:standalone}, with the the {\em network} confusion ratio $\rho$ (memory-aware social learning) being replaced by the {\em individual} confusion ratio $\rho_k$ (standalone learning). Despite this symmetry, the presence of the network indicator $\rho$ in place of the individual indicator $\rho_k$ makes a significant difference, as we now explain.


\begin{figure*}[t!]
	\centering
	\includegraphics[draft=false,width=.2\textwidth]{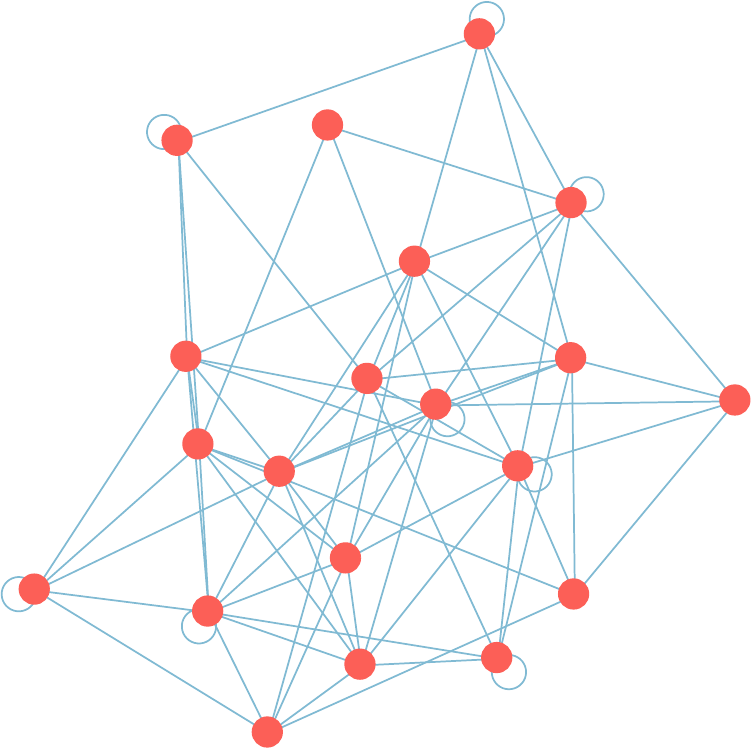}
	\includegraphics[draft=false,width=.26\textwidth]{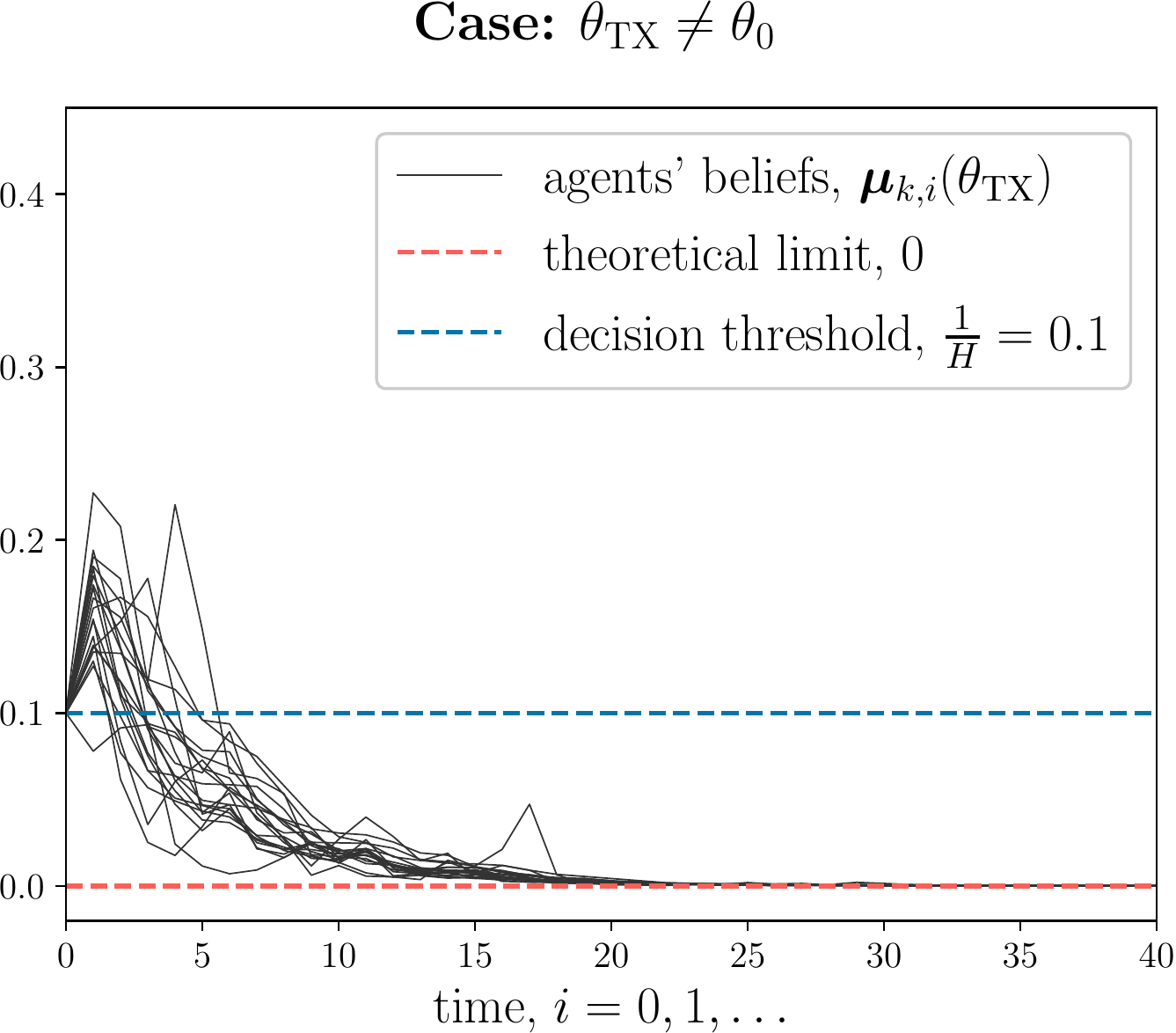}
	\includegraphics[draft=false,width=.26\textwidth]{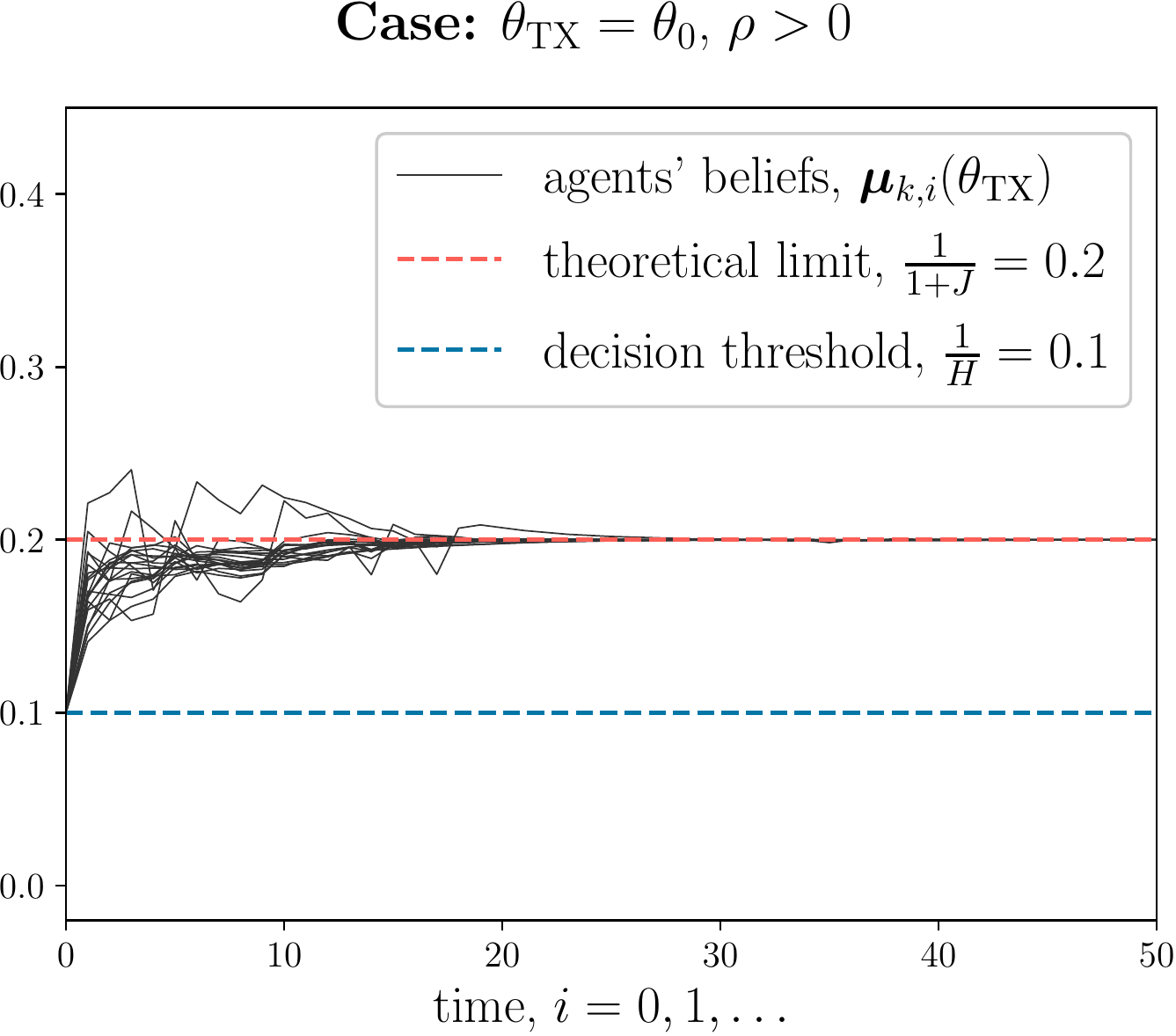}
	\includegraphics[draft=false,width=.26\textwidth]{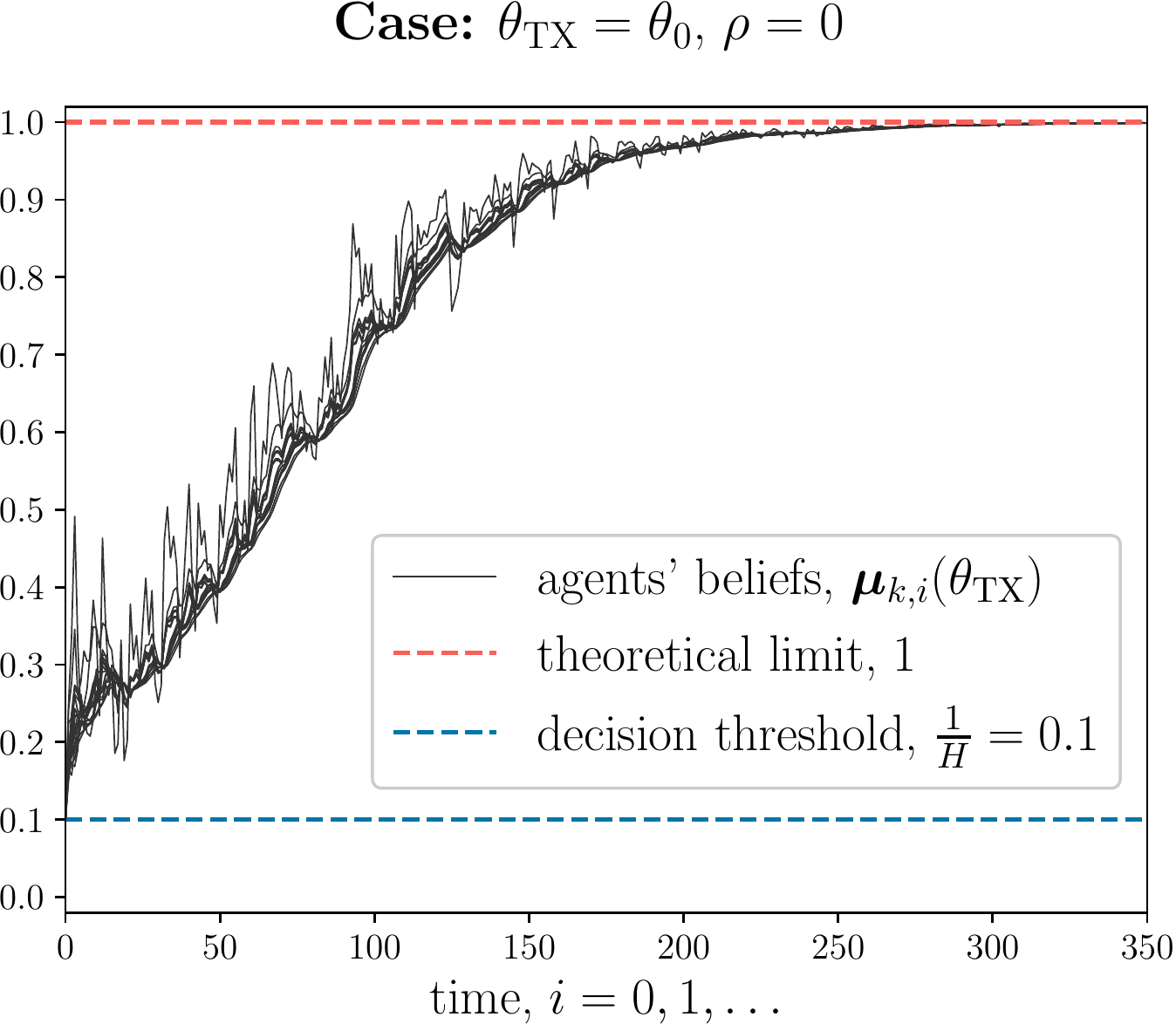}
	\caption{Dynamics of the beliefs of the transmitted hypothesis, $\bm{\mu}(\thetashare)$. We have $H=10$ hypotheses and the network of $K=20$ agents shown in the first panel. The agents are initialized with uniform beliefs $\mu_{k,0}(\theta)=1/H$. \textit{Second and third panels.} The algorithm is run with indistinguishable sets $\{\mathcal{I}_\ell\}_{\ell=1}^K$ provided in~\eqref{eq:III}; therefore, global identifiability holds. \textit{Fourth panel.} Case $\rho=\indcard=0$ enforced by letting $\mathcal{I}_1=\emptyset$.}
	\label{fig:TX}
\end{figure*}

One fundamental distinction between the standalone and the social learning algorithms resides in the way they treat indistinguishable hypotheses. 
In the standalone algorithm, we see that the true hypothesis and the indistinguishable hypotheses are in general assigned distinct beliefs. 
Since from the perspective of the standalone agent, the data does not convey useful information to discriminate between $\thetatrue$ and the indistinguishable hypotheses, distinct beliefs are caused by the initial belief. In other words, if the prior convictions bias the agent in favor of some hypothesis, this bias persists even in the long term. 
Consistently with this view, if the prior belief is uniform, from \eqref{eq:const-lim-standalone}, \eqref{eq:residualindistinguish-standalone}, and \eqref{eq:1-c-standalone} we conclude that the belief of $\thetatrue$ and the belief of the indistinguishable hypotheses converge to the same value $1/(1+\indcard_k)$, i.e., indistinguishability persists in the long term. No matter which decision rule is used, the standalone algorithm will be unable to discern.

The situation is significantly different for the social learning algorithm. 
Theorem~\ref{th:main-TXneq0} revealed that, when $\thetashare\neq\thetatrue$, cooperation leads each agent $k$ to correctly discard the transmitted hypothesis, irrespective of the initial beliefs, and even when $\thetashare$ would have been indistinguishable for a standalone agent $k$. 
We saw in Sec.~\ref{sec:decisionrule} that the capability of rejecting the indistinguishable hypotheses enables the design of a decision rule that achieves correct learning.  

Moreover, Theorem~\ref{th:standalone} reveals that agent $k$ in isolation can place full mass on $\thetatrue$ and zero mass on a hypothesis $\theta\neq\thetatrue$ only if $\rho_k=0$, i.e., if the problem is {\em locally} identifiable at agent $k$. 
In contrast, we learned from the previous analysis that an agent $k$ with a locally unidentifiable problem (i.e., $\rho_k\neq 0$) can place full mass on $\thetashare=\thetatrue$ by cooperating with a {\em single} agent that experiences local identifiability.

	\subsection{Comparison Against the Memoryless Strategy} \label{sec:comparison}
Now, we draw a parallel between the learning behavior achieved by the memoryless strategy~\eqref{eq:partialagnost} proposed in~\cite{bib:bordignon2020social} and the memory-aware strategy~\eqref{eq:model-memorypsi2}.
	It was shown in~\cite{bib:bordignon2020social} that with the memoryless strategy, the social learning problem can be interpreted as a binary detection problem involving the comparison between the likelihood of the true hypothesis, $L_k(\xi|\thetatrue)$ and a fictitious distribution: 
	\begin{equation}
	f_{k}(\xi)=\frac{1}{H-1}\sum_{\tau\neq\thetashare}L_k(\xi|\tau).
	\end{equation}
Then, the two main ingredients to ascertain the learning behavior are the following KL divergences:
	\begin{align}
d_{\rm TX}&\triangleq \sum_{k=1}^K v_k \,D_{\rm KL} (L_{k,\thetatrue}\|L_{k,\thetashare}),\\
d_{\mathrm{fict}}&\triangleq \sum_{k=1}^K v_k \,D_{\rm KL}(  L_{k,\thetatrue}\|f_{k} ).
\end{align}
%
%
Our analysis revealed that the memory-aware strategy is always able to classify correctly the hypothesis of interest.
This is not the case for the memoryless strategy. 
In particular, when $\thetashare=\theta_{0}$, the memoryless strategy always learns well~\cite{bib:bordignon2020social}. 
In contrast, when $\thetashare\neq\theta_{0}$, the memoryless strategy requires the condition $d_{\rm TX}>d_{\rm fict}$ to achieve correct learning~\cite[Theorem 3]{bib:bordignon2020social}. This means that, to discard $\thetashare$, the true distribution must be more similar to the fictitious distribution than to the distribution of the shared hypothesis.
In addition, the memoryless strategy exhibits an undesirable behavior when $d_{\rm TX}<d_{\rm fict}$, since the belief of $\thetashare$ converges to $1$, i.e., the strategy is completely fooled and ends up placing full mass on the wrong hypothesis~\cite{bib:bordignon2020social}.

\section{Simulation Results}
%

In this experimental section we consider the following setup. We generate a strongly-connected network with $K=20$ agents according to the Erd\H{o}s-R\'enyi model with connection probability $p=1/3$, and consider the Metropolis combination matrix~\cite{sayedbook}:
\begin{equation}
		a_{\ell k}=
		\begin{cases}
			1/\max\{|\mathcal{N}_{\ell}|,|\mathcal{N}_k|\}, &\ell\in\mathcal{N}_k\setminus\{k\},
			\\
			a_{\ell k}=0, &\ell\notin \mathcal{N}_k,
		\label{eq:combmatrix}
\end{cases}
\end{equation}
with $a_{kk}=1-\sum_{\ell\in\mathcal{N}_k}a_{\ell k}$. Note that the Perron eigenvector is uniform since $A$ is doubly stochastic.
The generated network is shown in the first panel of Fig.~\ref{fig:TX}. The choice in~\eqref{eq:combmatrix} yields a doubly-stochastic combination matrix, which has uniform Perron eigenvector, i.e., $v_k=1/K$ for all $k$.
We run on top of this network the social learning algorithm in~\eqref{eq:model-psi}--\eqref{eq:model-mu} with strategy~\eqref{eq:model-memorypsi2} and uniform initial beliefs $\mu_{k,0}(\theta)=1/H$,
and with:
\begin{equation}
	H=10,\qquad
	\Theta=\{1,2,\ldots,H\}, \qquad \thetatrue=1.
\end{equation}
For each agent $k$, we let $L_{k}(\xi|\theta)$ be the Gaussian distribution with unit variance and mean equal to $\thetatrue$ if $\theta\in\{\thetatrue\}\cup\mathcal{I}_k$, and equal to $\theta$ otherwise. We recall that the input measurements $\{\bm{\xi}_{k,i}\}_{i\geq1}$ of agent $k$ are generated according to $L_k(\xi|\thetatrue)$.


Moreover, we randomly generate the set of indistinguishable hypotheses $\{\mathcal{I}_k\}_{k=1}^K$
with a rule guaranteeing:
\begin{equation}
	|\mathcal{I}_k|=\left\{
	\begin{array}{ll}
		4, & \text{for $k\in[1,10]$,} \\
		8, & \text{for $k\in[11,15]$,} \\
		2, & \text{for $k\in[16,20]$,} \\
	\end{array}
	\right.
	\quad\text{and}\quad
	\bigcap_{k=1}^K \mathcal{I}_k=\emptyset.
	\label{eq:III}
\end{equation}
These constraints imply that global identifiability is satisfied.
Moreover, since the Perron eigenvector has equal entries $v_k=1/20$, we have:
	\begin{equation}
		\indcard=\Big(4^{10} \times 2^{5} \times 8^{5}\Big)^{1/20}=4.		
		\label{eq:I-ave}
	\end{equation}
In order to illustrate Theorems~\ref{th:main-TXneq0} and~\ref{th:main-TXeq0}, we consider the following experiments. 
The first experiment is run with $\thetashare\neq\thetatrue$, and the second panel of Fig.~\ref{fig:TX} illustrates the dynamics of the beliefs of agents relative to the transmitted hypothesis. We see that, according to~\eqref{eq:TXlim}, the beliefs all converge to $0$ due to the global identifiability property guaranteed by~\eqref{eq:III}.

In the third panel of Fig.~\ref{fig:TX} we illustrate the results for the experiment when $\thetashare=\thetatrue$. We see that the beliefs of all agents for $\thetashare$, as predicted by~\eqref{eq:const-lim-coroll}, and recalling~\eqref{eq:I-ave}, is such that:
\begin{equation}
	\bm{\mu}_{k,i}(\thetatrue)\rightarrow\frac{1}{1+\indcard}=\frac{1}{5}.
\end{equation}
We also consider a variation of the previous experiment where we enforce the condition $\rho=0$ by letting $\mathcal{I}_1=\emptyset$. 
We run the social learning algorithm for $\thetashare=\thetatrue$ and, as shown in the fourth panel of Fig.~\ref{fig:TX}, we see that the beliefs of all agents converge to the truth, i.e., $\bm{\mu}_{k,i}(\thetatrue)\rightarrow1$ for each $k$, as predicted by~\eqref{eq:const-lim}.

Finally, in the second and third panels of Fig.~\ref{fig:TX} we also draw with dashed blue line the threshold $1/H$ (we neglect the small $\varepsilon$ in \eqref{eq:fracHminusepsi}) that should be used by the decision rule proposed in Sec.~\ref{sec:decisionrule}. We see that the theoretical analysis is confirmed since: $i)$ when $\thetashare\neq\thetatrue$, as $i$ increases the beliefs fall below the threshold and the transmitted hypothesis is correctly discarded; and $ii)$ when $\thetashare=\thetatrue$, the beliefs end up staying above the threshold and the transmitted hypothesis is correctly accepted.   

\section{Conclusion}
We have examined the problem of social learning when agents share information only about a hypothesis of interest ({\em partial information sharing}). 
The main novelty introduced in this work is a {\em memory-aware} filling strategy, which each agent implements to turn the partial information received from its neighbors into a valid belief, to be used in the middle of the social learning updates. 
The analysis revealed the fundamental learning mechanism of memory-aware social learning with partial information sharing. 
Under the standard mild requirement of global identifiability, we established that there exists a decision rule that allows each agent to classify correctly the hypothesis of interest, with probability tending to $1$ as the number of iterations grows.
Notably, the standard maximum-belief rule is shown to fail in general under partial information sharing. 
Finally, the obtained results highlight the benefits of the memory-aware strategy over previous implementations.
The bottom line is that, under partial information sharing, the insertion of memory into the social learning mechanism is critical for a correct classification of the hypothesis of interest. 

\appendices 
\section{Preliminary Results}\label{sec:lemmata}

\begin{lemma}[Useful Submartingales] \label{th:sumart-true}
	Let Assumptions~\ref{th:finiteKL} and~\ref{th:positivemu0} hold. Let $\mathcal{S}_k$ be any nonempty agent-dependent set of hypotheses such that:
	\begin{equation}
		\mathcal{S}_k\subseteq \{\{\thetatrue\}\cup\mathcal{I}_k\}\setminus\{\thetashare\},
		\label{eq:Sk}
	\end{equation}
	and let $\mathcal{S}\triangleq\{\mathcal{S}_1,\mathcal{S}_2,\ldots,\mathcal{S}_K\}$. Define the nonpositive sequences:
	\begin{equation}
		\bm{m}_i \triangleq \sum_{k = 1}^K v_k \log \bm{\mu}_{k, i}
		(\thetatrue), \quad \bm{n}_i(\mathcal{S}) \triangleq \sum_{k = 1}^K v_k \log \bm{\mu}_{k, i}
		(\mathcal{S}_k),
		\label{eq:mndef}
	\end{equation}
	and:
	\begin{equation}
		\delta_k (\alpha) \triangleq D_{\rm KL} \Big( L_{k,\thetatrue}\Big\|\sum_{\tau\in\Theta}\alpha(\tau) L_{k,\tau}\Big),
		\label{eq:DKL-alpha}
	\end{equation}
	where $\alpha$ is a convex combination vector with dimension $H$. The following properties hold for any choice of $\thetashare$.
	\begin{enumerate}
		\item The random sequences
		$\bm{m}_i$ and $\bm{n}_i(\mathcal{S})$ fulfill:
		\begin{align}
			\mathbb{E} [\bm{m}_i |\mathcal{F}_{i - 1}] &\geq \bm{m}_{i - 1} + \sum_{k = 1}^K v_k \delta_k (\bm{\mu}_{k, i - 1}), \\
			\mathbb{E} [\bm{n}_i(\mathcal{S}) |\mathcal{F}_{i - 1}] &\geq \bm{n}_{i - 1}(\mathcal{S}) +
			\sum_{k = 1}^K v_k \delta_k (\bm{\mu}_{k, i - 1}),
		\end{align}
		where $\{\mathcal{F}_{i}\}_{i=0}^\infty$ is the filtration generated by the (deterministic) initial belief vector and by the input vector sequence $\{\bm{\xi}_{k,i}\}_{k=1}^K$, for $i\geq1$.
		\item Both $\bm{m}_i$ and $\bm{n}_i(\mathcal{S})$ are nonpositive submartingales with respect to~$\{\mathcal{F}_{i}\}_{i=0}^\infty$ and there exist random variables $\bm{m}_\infty$ and $\bm{n}_\infty(\mathcal{S})$ such that:
		\begin{equation}
			\bm{m}_i \xrightarrow{{\rm a.s.}} \bm{m}_\infty,
			\qquad
			\bm{n}_i(\mathcal{S}) \xrightarrow{\rm a.s.} \bm{n}_\infty(\mathcal{S}).
		\end{equation}
		\item The expectations $\mathbb{E} [\bm{m}_i]$ and $\mathbb{E} [\bm{n}_i(\mathcal{S})]$ have finite
		limit.
	\end{enumerate}
\end{lemma}

\begin{IEEEproof}
We first prove the claim for $\bm{m}_i$. Applying the arithmetic/geometric mean inequality we can write~\cite{bib:bullen}:
\begin{equation}
\sum_{\tau \in \Theta} \prod_{\ell = 1}^K 
\left[\widehat{\bm{\psi}}_{\ell k,i} (\tau)\right]^{a_{\ell k}}
\leq
\sum_{\tau \in \Theta} \sum_{\ell = 1}^K		
a_{\ell k}\widehat{\bm{\psi}}_{\ell k,i} (\tau)
=1.
\label{eq:leq1}
\end{equation}
	Therefore, in view of~\eqref{eq:model-mu} and~\eqref{eq:leq1} we get:
	\begin{equation}
		\bm{\mu}_{k, i} (\thetatrue) = \frac{\displaystyle{\prod_{\ell = 1}^K}
			\left[\widehat{\bm{\psi}}_{\ell k, i} (\thetatrue)\right]^{a_{\ell k}}}{\displaystyle{\sum_{\tau
					\in \Theta}} \, \displaystyle{\prod_{\ell = 1}^K} \left[\widehat{\bm{\psi}}_{\ell k, i}
			(\tau)\right]^{a_{\ell k}}} \geq \prod_{\ell = 1}^K
		\left[\widehat{\bm{\psi}}_{\ell k, i} (\thetatrue)\right]^{a_{\ell k}} , \label{eq:numdenom}
	\end{equation}
	which, using the definition of $\widehat{\bm{\psi}}_{\ell k, i}$ from~\eqref{eq:model-memorypsi2}, yields:
	\begin{equation}
		\bm{\mu}_{k, i} (\thetatrue) 
		\geq \left\{ \begin{array}{ll}
			\displaystyle{\prod_{\ell = 1}^K}
			\left[\bm{\psi}_{\ell, i} (\thetatrue)\right]^{a_{\ell k}}, 
			& \text{if $\thetashare=\thetatrue$}, \\
			\bm{\psi}_{k, i} (\thetatrue)\displaystyle{\frac{\displaystyle{\prod_{\ell = 1}^K} \left[1 \!-\!\bm{\psi}_{\ell, i}
					(\thetashare)\right]^{a_{\ell k}}}{1 - \bm{\psi}_{k, i}
					(\thetashare)}} ,
			& \text{if $\thetashare\neq\thetatrue$}.
		\end{array} \right.
		\label{eq:num2}
	\end{equation}
	Exploiting the identity $\sum_k a_{\ell k} v_k = v_{\ell}$, from~\eqref{eq:num2} we get, independently of the choice of $\thetashare$:
	\begin{equation}
		\bm{m}_i=\sum_{k = 1}^K v_k \log \bm{\mu}_{k, i} (\thetatrue) \geq \sum_{k =
			1}^K v_k \log \bm{\psi}_{k, i} (\thetatrue),
		\label{eq:martineq}
	\end{equation}
	which, further using~\eqref{eq:model-psi}, implies:
	\begin{equation}
		\bm{m}_i \geq \bm{m}_{i-1} +\sum_{k=1}^Kv_{ k}\log \frac{L_k(\bm{\xi}_{k,i}|\thetatrue)}{\displaystyle\sum_{\tau\in\Theta}L_k(\bm{\xi}_{k,i}|\tau)\bm{\mu}_{k,i-1}(\tau)}.
		\label{eq:slogslog}
	\end{equation}
	Now, taking the conditional expectation $\mathbb{E}\left[\,\cdot\,|\mathcal{F}_{i-1} \right]$ on both sides of~\eqref{eq:slogslog}, we obtain:
	\begin{equation}
		\mathbb{E}\left[\bm{m}_i|\mathcal{F}_{i-1} \right]\geq \bm{m}_{i-1} +\sum_{k=1}^Kv_k\delta_k(\bm{\mu}_{k,i-1})
		,\label{eq:aux1}
	\end{equation}
	where $\delta(\cdot)$ is defined in~\eqref{eq:DKL-alpha}. 
	This proves part $1)$ for $\bm{m}_i$.
	Since $\delta_k(\mu_{k,i-1})\geq0$ (it is a KL divergence), we also have that
	$\bm{m}_i$ is a submartingale:
	\begin{equation}
		\mathbb{E}\left[\bm{m}_{i}|\mathcal{F}_{i-1} \right]\geq \bm{m}_{i-1}.
		\label{eq:aux4}
	\end{equation}
	Therefore, calling upon the martingale convergence theorem~\cite{bib:billingsley}, the sequence $\bm{m}_i$ converges almost surely. 
	This proves part $2)$ for $\bm{m}_i$. 
	Finally, part $3)$ for $\bm{m}_i$ follows by taking the total expectation on both sides of~\eqref{eq:aux4}:
	\begin{equation}
		0\geq \mathbb{E}\left[\bm{m}_{i} \right]\geq \mathbb{E}\left[\bm{m}_{i-1} \right]\geq\dots \geq m_0,
		\label{eq:expineq}
	\end{equation}
	which implies that the sequence of expectations converges (since it is nondecreasing and bounded from above).
	
	Now we focus on $\bm{n}_i(\mathcal{S})$. In view of~\eqref{eq:model-mu},~\eqref{eq:subsetS} and~\eqref{eq:leq1}:
	\begin{equation}
		\bm{\mu}_{k, i} (\mathcal{S}_k)\! = \!\! \sum_{\theta \in \mathcal{S}_k}\!
		\frac{ \displaystyle{\prod_{\ell = 1}^K} \left[\widehat{\bm{\psi}}_{\ell k, i}
			(\theta)\right]^{a_{\ell k}}}{\displaystyle{\sum_{\tau \in \Theta} \prod_{\ell = 1}^K} \!
			\left[\widehat{\bm{\psi}}_{\ell k, i} (\tau)\right]^{a_{\ell k}}} 
		\!\geq\!\!
		\sum_{\theta \in \mathcal{S}_k}\!
		\displaystyle{\prod_{\ell = 1}^K} \left[\widehat{\bm{\psi}}_{\ell k, i}
		(\theta)\right]^{a_{\ell k}}
		\label{eq:defmu}
	\end{equation}
	and since from the definition of $\mathcal{S}_k$ in \eqref{eq:Sk} we see that $\thetashare \notin \mathcal{S}_k$ for any agent $k$, by applying~\eqref{eq:model-memorypsi2} and~\eqref{eq:subsetS} we obtain:
	\begin{equation}
		\bm{\mu}_{k, i} (\mathcal{S}_k) \geq  \bm{\psi}_{k, i} (\mathcal{S}_k)\frac{\displaystyle{\prod_{\ell = 1}^K} \big[1 -\bm{\psi}_{\ell, i}
			(\thetashare)\big]^{a_{\ell k}}}{1 -\bm{\psi}_{k, i}
			(\thetashare)},
		\label{eq:mu-Theta-heq}
	\end{equation}
	which, using the identity $\sum_k a_{\ell k} v_k = v_{\ell}$, implies the inequality:
	\begin{align}
		\bm{n}_i(\mathcal{S}) &\geq
		\sum_{k = 1}^K v_k \log \bm{\psi}_{k, i} (\mathcal{S}_k)
		\nonumber\\
		&=\bm{n}_{i-1}(\mathcal{S})+\sum_{k=1}^Kv_{ k}\log \frac{L_k(\bm{\xi}_{k,i}|\thetatrue)}{\displaystyle\sum_{\tau\in\Theta}L_k(\bm{\xi}_{k,i}|\tau)\bm{\mu}_{k,i-1}(\tau)},
	\end{align}
	where the equality follows from \eqref{eq:model-psi} and the fact that for $\theta\in\mathcal{S}_k$ we have $L_{k,\theta}=L_{k,\thetatrue}$. 
	The proof can be completed by repeating the same steps made to prove \eqref{eq:aux1}--\eqref{eq:expineq} starting from \eqref{eq:slogslog}.
\end{IEEEproof}

\begin{corollary}[Expectation of Log-Beliefs] \label{th:psineq}
	Let Assumptions~\ref{th:finiteKL} and~\ref{th:positivemu0} hold. For all $k=1,2,\ldots,K$ and all $i \geq 1$ we have:
	\begin{equation}
		\mathbb{E}\left[ \log \frac{1}{\bm{\psi}_{k,i}(\thetatrue)}\right] \leq \frac{|m_0|}{v_k}, \quad \mathbb{E}\left[ \log \frac{1}{\bm{\psi}_{k,i}(\mathcal{S}_k)}\right] \leq \frac{|n_0(\mathcal{S})|}{v_k}.
	\end{equation}
\end{corollary}
\begin{IEEEproof}
	Let us first prove the claim for $\bm{m}_i$. Using~\eqref{eq:model-psi} and~\eqref{eq:DKL-alpha}, we have that:
	\begin{align}
		\mathbb{E}\left[\log \frac{1}{\bm{\psi}_{k,i}(\thetatrue)}\right]
		&=
		\mathbb{E}\left[\log\frac{1}{\bm{\mu}_{k,i-1}(\thetatrue)}\right]
		- \mathbb{E}\left[\delta_k(\bm{\mu}_{k,i-1})\right]
		\nonumber\\
		&\leq\mathbb{E}\left[\log \frac{1}{\bm{\mu}_{k,i-1}(\thetatrue)}\right],\label{eq:aux5}
	\end{align}
	where the inequality follows from the fact that $\delta_k(\bm{\mu}_{k,i})$ is nonnegative (it is a KL divergence).
	Noting that by definition $v_k\log\bm{\mu}_{k,i-1}(\thetatrue) \geq \bm{m}_{i-1}$, and using~\eqref{eq:expineq}, we have:
	\begin{equation}
		\mathbb{E}\left[\log\frac{1}{\bm{\mu}_{k,i-1}(\thetatrue)}\right] \leq - \frac{\mathbb{E}[\bm{m}_{i-1}]}{v_k} \leq - \frac{m_0}{v_k} = \frac{|m_0|}{v_k},
		\label{eq:aboveineq}
	\end{equation}
	and by replacing~\eqref{eq:aboveineq} into~\eqref{eq:aux5} we get the claim for $\bm{m}_i$. 
	Now, using~\eqref{eq:model-psi} and~\eqref{eq:subsetS}, and noting that $L_{k,\theta}=L_{k,\thetatrue}$ for $\theta\in\mathcal{S}_k$, we have:
	\begin{align}
		\mathbb{E}\left[\log \frac{1}{\bm{\psi}_{k,i}(\mathcal{S}_k)}\right]
		&=
		\mathbb{E}\left[\log\frac{1}{\bm{\mu}_{k,i-1}(\mathcal{S}_k)}\right]
		- \mathbb{E}[\delta_k(\bm{\mu}_{k,i-1})].
		\label{eq:aux5-2}
	\end{align}
	The proof can be completed by repeating the same steps made to obtain~\eqref{eq:aboveineq} from~\eqref{eq:aux5}.
\end{IEEEproof}

\begin{lemma}[All Agents Discard Distinguishable Hypotheses] \label{th:dist0}
	Let $\mathcal{D}_k\neq\emptyset$ and let Assumptions~\ref{th:finiteKL},~\ref{th:positivemu0} and~\ref{th:convexcomb} hold. 
	For all $k=1,2,\ldots,K$ and all $\tau\in\mathcal{D}_k$:
	\begin{equation}
		\bm{\mu}_{k,i}(\tau) \xrightarrow{\rm p} 0.
		\label{eq:distishable0}
	\end{equation}
\end{lemma}
\begin{IEEEproof}
	We prove the claim by considering an arbitrary agent $k$ with $|\mathcal{D}_k|>0$.
	Since from~\eqref{eq:aux1} we have:
	\begin{equation}
		0\leq
		\sum_{k=1}^K v_k\E \left[\delta_k(\bm{\mu}_{k,i-1})\right]
		\leq
		\E[\bm{m}_i]- \E[\bm{m}_{i-1}].
	\end{equation} 
	Then, in view of part $3)$ of Lemma~\ref{th:sumart-true}, and using the squeeze theorem, we have:
	\begin{equation}
		\lim_{i\rightarrow\infty}
		\sum_{k=1}^K v_k\E \left[\delta_k(\bm{\mu}_{k,i-1})\right]=0.
	\end{equation}
	Recalling that $v_k>0$, we conclude that $\delta_k(\bm{\mu}_{k,i-1})$ converges to zero in mean, and therefore also in probability:
	\begin{equation}
		\delta_k(\bm{\mu}_{k,i-1}) \xrightarrow{\rm p} 0.
		\label{eq:KLgoestozero}
	\end{equation}
Introducing the total variation distance $\|f - g\|$ between the probability measures associated to two distributions $f$ and $g$, and applying Pinsker's inequality~\cite{cover1999elements}, we can lower bound the KL divergence $\delta_k(\bm{\mu}_{k,i-1})$ as follows:
	\begin{align}
		\delta_k(\bm{\mu}_{k,i-1})\! &\geq
		\frac 1 2\Big\| L_k(\thetatrue) - \sum_{\tau\in\Theta}\bm{\mu}_{k,i-1}(\tau) L_k(\tau)\Big\|^2 \nonumber \\
		& = \frac{1}{2}
		\Big(\sum_{\theta\in\mathcal{D}_{k}}\!\!\bm{\mu}_{k,i-1}(\theta)\!\Big)^{\!\!2}
		\Big\|
		L_{k}(\thetatrue) \!-\!\!\!
		\sum_{\tau\in\mathcal{D}_{k}}\!\!\!
		\bm{\alpha}(\tau)L_{k}(\tau)
		\Big\|^2, 
		\label{eq:abc}
	\end{align}
where the equality holds since:
	\begin{align}
		&L_{k}(\thetatrue) - \sum_{\tau\in\Theta}\bm{\mu}_{k,i-1}(\tau) L_k(\tau)\nonumber\\
		&=
		\Bigg(1-\!\!\sum_{\tau\in\mathcal{I}_k\cup\{\thetatrue\}}\bm{\mu}_{k,i-1}(\tau)\Bigg)L_{k}(\thetatrue )-\!\!\sum_{\tau\in\mathcal{D}_{k}}\!\bm{\mu}_{k,i-1}(\tau) L_{k}(\tau)\nonumber\\
		&=
		\sum_{\tau\in\mathcal{D}_{k}}\bm{\mu}_{k,i-1}(\tau)
		\Bigg(
		L_{k}(\thetatrue) - 
		\sum_{\tau\in\mathcal{D}_{k}}
		\bm{\alpha}(\tau)L_{k}(\tau)
		\Bigg),
		\label{eq:convexcomb}
	\end{align}
	having defined $\bm{\alpha}(\tau)=\frac{\bm{\mu}_{k,i-1}(\tau)}{\sum\limits_{\theta\in\mathcal{D}_{k}}\bm{\mu}_{k,i-1}(\theta)}$.
Observe now that Assumption~\ref{th:convexcomb} requires that 
the true likelihood is different from {\em any} convex combination of the distinguishable hypotheses. 
Since it can be shown that the total variation distance: 
	\begin{equation}
		\Big\|
		L_{k}(\thetatrue) - 
		\sum_{\tau\in\mathcal{D}_{k}}
		\alpha(\tau)L_{k}(\tau)
		\Big\|
		\label{eq:totalv}
	\end{equation}
is continuous with respect to $\alpha$, we conclude that the minimum value $b$ of this distance is taken at some point(s) $\alpha$, and this minimum must be strictly positive, otherwise we would have an $\alpha$ for which $L_{k}(\thetatrue)$ could be written as a convex combination of the distinguishable likelihoods, violating Assumption~\ref{th:convexcomb}. Therefore, we can write:
\begin{equation}
		\Big\|
		L_{k}(\thetatrue) - 
		\sum_{\tau\in\mathcal{D}_{k}}
		\alpha(\tau)L_{k}(\tau)
		\Big\|\geq b>0,
\end{equation}
which, by substituting into \eqref{eq:abc}, yields
	\begin{equation}
	\delta_k(\bm{\mu}_{k,i-1})
	\geq
	\frac{b^2}{2}
	\left(\sum_{\theta\in\mathcal{D}_{k}}\!\!\bm{\mu}_{k,i-1}(\theta)\!\right)^{\!\!2},
	\end{equation}
and the proof is complete in view of \eqref{eq:KLgoestozero}.
\end{IEEEproof}

\begin{lemma}[Evolution of Non-Transmitted Hypotheses]\label{lem:ntxhyp}
	Let Assumptions~\ref{th:finiteKL} and~\ref{th:positivemu0} hold. Assume that $ \{\{\thetatrue\}\cup\mathcal{I}_k\}\setminus\{\thetashare\}$ is nonempty.
For any $\tau \in \mathcal{D}_k\setminus\{\thetashare\}$:
	\begin{equation}
		\bm{\mu}_{k,i}(\tau)\xrightarrow{\rm a.s.}
		0.
	\end{equation}
\end{lemma}
\begin{IEEEproof}
	Let us consider $\tau\in\mathcal{D}_k\setminus\{\thetashare\}$ and $\theta\in\{\{\thetatrue\}\cup\mathcal{I}_k\}\setminus\{\thetashare\}$. Using~\eqref{eq:model-psi},~\eqref{eq:model-mu}, and~\eqref{eq:model-memorypsi2}, we can
	write the recursion:
	\begin{equation}
		\log\frac{\bm{\mu}_{k,i}(\theta)}{\bm{\mu}_{k,i}(\tau)}=\log\frac{\bm{\mu}_{k,i-1}(\theta)}{\bm{\mu}_{k,i-1}(\tau)}+\log \frac{L_k(\bm{\xi}_{k,i}|\thetatrue)}{L_k(\bm{\xi}_{k,i}|\tau)}
	\end{equation}
	and unfolding it we get:
	\begin{equation}
		\log\frac{\bm{\mu}_{k,i}(\theta)}{\bm{\mu}_{k,i}(\tau)}=\log \frac{\mu_{k,0}(\theta)}{\mu_{k,0}(\tau)}+\sum_{j=1}^i\log \frac{L_k(\bm{\xi}_{k,j}|\thetatrue)}{L_k(\bm{\xi}_{k,j}|\tau)}.\label{eq:lemmantx}
	\end{equation}
	Now divide~\eqref{eq:lemmantx} by $i$ and take the limit as $i$ goes to infinity. In view of Assumptions~\ref{th:finiteKL} and~\ref{th:positivemu0}, and since $\tau\in\mathcal{D}_k$, the strong law of large numbers yields to the finite, strictly positive limit:
	\begin{equation}
		\frac{1}{i}\log\frac{\bm{\mu}_{k,i}(\theta)}{\bm{\mu}_{k,i}(\tau)}
		\xrightarrow{\rm a.s.} D_{\rm KL}(L_{k,\thetatrue}\|L_{k,\tau})
		,\label{eq:lemmantx1}
	\end{equation}
	and therefore $\log\frac{\bm{\mu}_{k,i}(\theta)}{\bm{\mu}_{k,i}(\tau)}\xrightarrow{\rm a.s.}+\infty$, leading to the claim.
\end{IEEEproof}

\begin{lemma}[Ratios of Beliefs of Distinguishable and Indistinguishable Hypotheses] \label{th:mart00}
	Let $\thetashare=\theta_{0}$, and let Assumptions~\ref{th:finiteKL},~\ref{th:positivemu0} and~\ref{th:convexcomb} hold. For each $k$, the sequence $\frac{\bm{\psi}_{k, i} (\mathcal{D}_k)}{\bm{\psi}_{k, i} (\mathcal{I}_k)}$
is a nonnegative martingale with respect to the filtration $\{\mathcal{F}_{i}\}_{i=0}^\infty$ generated by the (deterministic) initial belief vector and by the input vector sequence $\{\bm{\xi}_{k,i}\}_{k=1}^K$ for $i\geq1$. Moreover, this martingale vanishes almost surely.
\end{lemma}
\begin{IEEEproof}
	In view of~\eqref{eq:model-psi} and~\eqref{eq:subsetS} we have:
	\begin{align}
		\frac{\bm{\psi}_{k, i} (\mathcal{D}_k)}{\bm{\psi}_{k, i}
			(\mathcal{I}_k)} & = \frac{\displaystyle{\sum_{\tau \in \mathcal{D}_k}} \bm{\psi}_{k,
				i} (\tau)}{\displaystyle{\sum_{\tau \in \mathcal{I}_k}} \bm{\psi}_{k, i} (\tau)}
		\nonumber\\
		& = \frac{1}{\bm{\mu}_{k, i - 1} (\mathcal{I}_k)} \sum_{\tau \in
			\mathcal{D}_k} \bm{\mu}_{k, i - 1} (\tau)
		\frac{L_k (\bm{\xi}_{k, i} | \tau)}{L_k
			(\bm{\xi}_{k, i} | \thetatrue)},
	\end{align}
	and since for each $\tau\in\Theta$:
	\begin{equation}
		\mathbb{E} \left[ \frac{L_k (\bm{\xi}_{k, i} | \tau)}{L_k
			(\bm{\xi}_{k, i} | \thetatrue)} \right] = 1,
	\end{equation}
	we conclude that the random sequence $\frac{\bm{\psi}_{k, i} (\mathcal{D}_k)}{\bm{\psi}_{k, i} (\mathcal{I}_k)}$ is a martingale:
	\begin{align}
		&\mathbb{E} \left[ \frac{\bm{\psi}_{k, i} (\mathcal{D}_k)}{\bm{\psi}_{k, i}
			(\mathcal{I}_k)} \middle|\mathcal{F}_{i - 1}
		\right] \nonumber \\& = \frac{1}{\mu_{k, i - 1} (\mathcal{I}_k)} \sum_{\tau \in
			\mathcal{D}_k} \mu_{k, i - 1} (\tau) \mathbb{E} \left[ \frac{L_k
			(\bm{\xi}_{k, i} | \tau)}{L_k (\bm{\xi}_{k, i} | \thetatrue)}
		\right] \nonumber\\
		& = \frac{\mu_{k, i - 1} (\mathcal{D}_k)}{\mu_{k, i - 1} (\mathcal{I}_k)} 
		= \frac{\psi_{k,i-1}(\mathcal{D}_k)}{\psi_{k,i-1}(\mathcal{I}_k)} \nonumber\\ &\quad\times \underbrace{\frac{\prod_{\ell=1}^K \left[1-\psi_{\ell,i-1}(\thetatrue)\right]^{a_{\ell k}}}{1-\psi_{k,i-1}(\thetatrue)}\frac{1-\psi_{k,i-1}(\thetatrue)}{\prod_{\ell=1}^K \left[1-\psi_{\ell,i-1}(\thetatrue)\right]^{a_{\ell k}}}}_{=1},
	\end{align}
	where the last equality follows from~\eqref{eq:model-psi} and~\eqref{eq:model-memorypsi2}. Therefore, the sequence $\frac{\bm{\psi}_{k, i} (\mathcal{D}_k)}{\bm{\psi}_{k, i} (\mathcal{I}_k)}$ converges almost surely~\cite{bib:billingsley}, and to prove that the limit is $0$ it suffices to show that:
	\begin{equation}
		\frac{\bm{\psi}_{k, i} (\mathcal{D}_k)}{\bm{\psi}_{k, i}
			(\mathcal{I}_k)} \xrightarrow{\rm p} 0.
		\label{eq:p-conv}
	\end{equation}
	The convergence in~\eqref{eq:p-conv} comes from~\cite[Lemma 7]{bib:bordignon2020social}, where {\em i)} the sequence $\bm{\psi}_{k, i} (\mathcal{D}_k)$ plays the role of $\bm{x}_i$, since from Lemma~\ref{th:dist0} and Assumption~\ref{th:finiteKL} we know that it converges to $0$ in probability; and {\em ii)} the sequence $\frac{1}{\bm{\psi}_{k, i}(\mathcal{I}_k)}$ plays the role of $\bm{y}_i$, since calling upon Markov's inequality and Corollary~\ref{th:psineq} for $\mathcal{S}_k=\mathcal{I}_k$ we have:
	\begin{align}
		&\mathbb{P} \left[ \frac{1}{\bm{\psi}_{k, i} (\mathcal{I}_k)} > M \right]  =
		\mathbb{P} \left[ \log \frac{1}{\bm{\psi}_{k, i} (\mathcal{I}_k)} >
		\log (M) \right] \nonumber\\
		& \leq \frac{1}{ \log (M)} \mathbb{E} \left[ \log
		\frac{1}{\bm{\psi}_{k, i} (\mathcal{I}_k)} \right] 
		\leq  \frac{1}{\log (M)} \frac{|n_0(\mathcal{S})|}{v_k}  \stackrel{M\rightarrow\infty}{\longrightarrow} 0.
		\label{eq:P<M}
	\end{align}
\end{IEEEproof}

\section{Proof of Theorem~\ref{th:main-TXneq0}}
\label{sec:main-TXneq0-proof}
Since $\thetashare\neq\theta_{0}$, then $ \{\{\thetatrue\}\cup\mathcal{I}_k\}\setminus\{\thetashare\}$ is nonempty and~\eqref{eq:0-again} follows directly from Lemma~\ref{lem:ntxhyp}. Therefore, since:
\begin{equation}
	\bm{\mu}_{k, i} (\thetashare)+ \bm{\mu}_{k, i} (\{\{\thetatrue\}\cup\mathcal{I}_k\}\setminus\{\thetashare\})+ \bm{\mu}_{k, i} (\mathcal{D}_k\setminus\{\thetashare\})\!=\!1,
	\label{eq:beliefsumtoone}
\end{equation}
and since~\eqref{eq:TXlim} is an immediate consequence of~\eqref{eq:0-again} and~\eqref{eq:residual}, it remains to prove~\eqref{eq:residual}.

Under global identifiability we have $\thetashare\in\mathcal{D}_h$ for some agent $h$, and therefore from Lemma~\ref{th:dist0} we have:
\begin{equation}
	\bm{\mu}_{h,i}(\thetashare)\xrightarrow{\rm p}0.
	\label{eq:claimTX}
\end{equation}
Now, define:
\begin{equation}
	\bm{u}_i \triangleq \frac{L_h(\bm{\xi}_{h,i}|\thetashare)}{L_h(\bm{\xi}_{h,i}|\thetatrue)},
\end{equation}
\begin{equation}
	\bm{v}_i \triangleq \frac{\bm{\mu}_{h,i-1}(\thetashare)}{\displaystyle{\sum_{\tau\in{\mathcal{D}_h}}} \frac{L_h(\bm{\xi}_{h,i}|\tau)}{L_h(\bm{\xi}_{h,i}|\thetatrue)} \bm{\mu}_{h,i-1}(\tau) + \Big(1- \bm{\mu}_{h,i-1}\big(
		\mathcal{D}_h
		\big) \Big)  }.
\end{equation}
Calling upon Slutsky's theorem we have~\cite{ShaoBook}:
\begin{equation}
	\bm{\psi}_{h,i}(\thetashare) 
	= \bm{u}_i \bm{v}_i \xrightarrow{\rm p} 0.
	\label{eq:psi0}
\end{equation}
In fact: {\em i)} term $\bm{u}_i$, which is well defined due to Assumption~\ref{th:finiteKL}, has constant distribution over time, whereas 
{\em ii)} term $\bm{v}_i$ vanishes in probability, as one can verify from~\eqref{eq:distishable0},~\eqref{eq:claimTX} and using Slutsky's theorem a second time to obtain:
\begin{equation}
	\sum_{\tau \in \mathcal{D}_{h}} \frac{L_{h} (\bm{\xi}_{h, i} | \tau)}{L_{h} (\bm{\xi}_{h, i} | \thetatrue)} \bm{\mu}_{h, i - 1} (\tau) \xrightarrow{\rm p}0.
\end{equation}
Consider now an agent $k$ for which $a_{hk}>0$. Using~\eqref{eq:model-mu} and~\eqref{eq:model-memorypsi2} we can write:
\begin{align}
	\bm{\mu}_{k,i}(\thetashare) &\leq
	\frac{\bm{\mu}_{k,i}(\thetashare)}{1-\bm{\mu}_{k,i}(\thetashare)} 
	= 
	\prod_{\ell=1}^K \left(\frac{\bm{\psi}_{\ell,i}(\thetashare)}{1-\bm{\psi}_{\ell,i}(\thetashare)}\right)^{a_{\ell k}}
	\nonumber\\ &\leq \ \left[\bm{\psi}_{h,i}(\thetashare)\right]^{a_{h k}}
	\prod_{\ell=1}^K \left(\frac{1}{\bm{\psi}_{\ell,i}(\thetatrue)}\right)^{a_{\ell k}},
\label{eq:ineqchainappendix22}
\end{align}
where we used the fact that:
\begin{equation}
	1-\bm{\psi}_{\ell,i}(\thetashare) \geq \bm{\psi}_{\ell,i}(\thetatrue)
\end{equation}
since $\thetashare\neq\thetatrue$.
Now we apply~\cite[Lemma 7]{bib:bordignon2020social} to the RHS of \eqref{eq:ineqchainappendix22} with: {\em i)} the sequence $\left[\bm{\psi}_{h,i}(\thetashare)\right]^{a_{h k}}$ in the role of $\bm{x}_i$, since it converges in probability to zero according to~\eqref{eq:psi0}; and {\em ii)} the sequence $\prod_{\ell = 1}^K \left( \frac{1}{\bm{\psi}_{\ell, i} (\thetatrue)} \right)^{a_{\ell k}}$ in the role of $\bm{y}_i$, since exploiting Markov's inequality and Corollary~\ref{th:psineq} we have:
\begin{align}
	\lefteqn{\mathbb{P} \left[ \prod_{\ell = 1}^K \left( \frac{1}{\bm{\psi}_{\ell, i} (\thetatrue)} \right)^{a_{\ell k}} > M \right]}
	\nonumber\\ & =
	\mathbb{P} \left[ \sum_{\ell=1}^K a_{\ell k} \log \frac{1}{\bm{\psi}_{\ell, i} (\thetatrue)} >
	\log M \right] \nonumber\\
	& \leq \frac{1}{ \log M} \sum_{\ell=1}^K a_{\ell k} \mathbb{E} \left[ \log
	\frac{1}{\bm{\psi}_{\ell, i} (\thetatrue)} \right] \nonumber\\
	& \leq \frac{|m_0|}{\log M} \sum_{\ell=1}^K \frac{a_{\ell k}}{v_\ell}  \stackrel{M\rightarrow\infty}{\longrightarrow} 0.
	\label{eq:prod-P<M}
\end{align}
Therefore, from~\cite[Lemma 7]{bib:bordignon2020social} we conclude that the RHS of \eqref{eq:ineqchainappendix22} vanishes in probability, implying that $\bm{\mu}_{k,i}(\thetashare)\xrightarrow{\rm p}0$.
Since the network is strongly connected (Assumption~\ref{th:strong}), by iterating the above reasoning we conclude that $\bm{\mu}_{k, i} (\thetashare) \xrightarrow{\rm p} 0$ for all agents, which, combined with~\eqref{eq:0-again} and~\eqref{eq:beliefsumtoone}, yields:
\begin{equation}
	\bm{\mu}_{k, i}(\{\{\thetatrue\}\cup\mathcal{I}_k\}\setminus\{\thetashare\})\xrightarrow{\rm p}1.
	\label{eq:skconvp}
\end{equation}
By further using~\eqref{eq:Sk} and~\eqref{eq:mndef} with the choice $\mathcal{S}_k=\{\{\thetatrue\}\cup\mathcal{I}_k\}\setminus\{\thetashare\}$, from \eqref{eq:skconvp} we get:
$\bm{n}_i(\mathcal{S}) \xrightarrow{\rm p}0$.
From part $2)$ of Lemma~\ref{th:sumart-true}, this convergence must take place almost surely: 
\begin{equation}
	\bm{n}_i(\mathcal{S})=\sum_{k=1}^Kv_k\log \bm{\mu}_{k,i}(\mathcal{S}_k) \xrightarrow{\rm a.s.}0,
\end{equation}
and since $v_k>0$ for all $k$, we must have:
\begin{equation}
	\bm{\mu}_{k,i}(\mathcal{S}_k)=\bm{\mu}_{k, i} (\mathcal{I}_k\setminus\{\thetashare\})+\bm{\mu}_{k, i} (\thetatrue) \xrightarrow{\rm a.s.}1,\label{eq:lim22}
\end{equation}
for every agent $k$.
Finally, since from~\eqref{eq:model-mu} we have that for any indistinguishable hypothesis $\theta \in \mathcal{I}_k\setminus\{\thetashare\}$:
\begin{equation} 
	\frac{\bm{\mu}_{k, i} (\theta)}{\bm{\mu}_{k, i} (\thetatrue)}
	= \frac{\mu_{k, 0} (\theta)}{\mu_{k, 0} (\thetatrue)},
	\label{eq:goldenratio}
\end{equation}
which implies \eqref{eq:residual} in view of \eqref{eq:subsetS}, and the proof is complete.

\section{Proof of Theorem~\ref{th:main-TXeq0}}
\label{sec:main-TXeq0-proof}

We consider separately the cases $\rho=0$ and $\rho>0$.

{\em I. Case $\rho=0$}.
First of all, note that for $\rho=0$ we must prove only \eqref{eq:const-lim}.
In view of \eqref{eq:confunet}, if $\rho=0$ we must have a ``strong'' agent $h$ with $\rho_h=0$, a condition that, in view of \eqref{eq:confuratio} and Assumption~\ref{th:positivemu0}, implies $\mathcal{I}_h=\emptyset$.
From Lemma~\ref{th:dist0} we have that:
\begin{equation}
	\bm{\mu}_{h,i}(\thetatrue)\xrightarrow{\rm p} 1.
	\label{eq:mustar1}
\end{equation}
Now using~\eqref{eq:model-psi} we  can write:
\begin{equation}
	\bm{\psi}_{h, i} (\thetatrue) 
	= \frac{\bm{\mu}_{h, i - 1} (\thetatrue)}{\displaystyle{\sum_{\tau \in \Theta \setminus \{\theta_{0}\}}} \frac{L_{h} (\bm{\xi}_{h, i} | \tau)}{L_{h} (\bm{\xi}_{h, i} | \thetatrue)} \bm{\mu}_{h, i - 1} (\tau) + \bm{\mu}_{h,i-1}(\thetatrue)}.
\end{equation}
Since {\em i)} the ratios $L_{h} (\bm{\xi}_{h, i} | \tau) / L_{h} (\bm{\xi}_{h, i} | \thetatrue)$ are identically distributed, and {\em ii)} $\bm{\mu}_{h,i-1}(\tau)$ vanishes in probability for $\tau\in\Theta\setminus\{\theta_{0}\}$ in view of~\eqref{eq:mustar1}, from Slutsky's theorem we have~\cite{ShaoBook}:
\begin{equation}
	\sum_{\tau \in \Theta\setminus\{\theta_{0}\}} \frac{L_{h} (\bm{\xi}_{h, i} | \tau)}{L_{h} (\bm{\xi}_{h, i} | \thetatrue)} \bm{\mu}_{h, i - 1} (\tau) \xrightarrow{\rm p}0,
	\label{eq:nec2}
\end{equation}
and therefore from~\eqref{eq:mustar1}--\eqref{eq:nec2} we get:
\begin{equation}
	\bm{\psi}_{h, i} (\thetatrue)\xrightarrow{\rm p} 1.
	\label{eq:psi0-1}
\end{equation}
Consider now an agent $k$ with $\rho_k>0$ and $a_{hk}>0$.
In view of~\eqref{eq:model-memorypsi2} we can write (recall that we are considering the case $\thetashare=\thetatrue$):
\begin{align}
	\bm{\mu}_{k, i} (\mathcal{I}_k)
	&<\frac{\bm{\mu}_{k, i} (\mathcal{I}_k)}{\bm{\mu}_{k, i} (\thetatrue)}
	= \frac{\bm{\psi}_{k, i} (\mathcal{I}_k)}{1 -\bm{\psi}_{k, i}
		(\thetatrue)} \prod_{\ell = 1}^K \left( \frac{1 -\bm{\psi}_{\ell, i}
		(\thetatrue)}{\bm{\psi}_{\ell, i} (\thetatrue)} \right)^{a_{\ell k}} 
	\nonumber\\&< (1 -\bm{\psi}_{h, i}
	(\thetatrue))^{a_{h k}} \prod_{\ell = 1}^K \left( \frac{1}{\bm{\psi}_{\ell, i} (\thetatrue)} \right)^{a_{\ell k}} \xrightarrow{\rm p} 0,
	\label{eq:prev}
\end{align}
where we used the fact that:
\begin{equation}
	\frac{\bm{\psi}_{k, i} (\mathcal{I}_k)}{1 -\bm{\psi}_{k, i}
		(\thetatrue)}=\frac{\bm{\psi}_{k, i} (\mathcal{I}_k)}{\bm{\psi}_{k, i}
		(\mathcal{I}_k) + \bm{\psi}_{k, i}
		(\mathcal{D}_k)}\leq\!1,
\end{equation}
whereas the convergence comes from~\cite[Lemma 7]{bib:bordignon2020social} where: {\em i)} the sequence $(1 -\bm{\psi}_{h, i}
(\thetatrue))^{a_{h k}}$ plays the role of $\bm{x}_i$, since it converges in probability to zero according to~\eqref{eq:psi0-1}; and {\em ii)} the sequence $\prod_{\ell = 1}^K \left( \frac{1}{\bm{\psi}_{\ell, i} (\thetatrue)} \right)^{a_{\ell k}}$ plays the role of $\bm{y}_i$, as it can be seen by exploiting Corollary~\ref{th:psineq} as we already did in~\eqref{eq:prod-P<M}. Thus, since $ \bm{\mu}_{k, i}
(\theta_{0}) = 1 - \bm{\mu}_{k, i} (\mathcal{I}_k) - \bm{\mu}_{k, i} (\mathcal{D}_k)$, from~\eqref{eq:distishable0} and~\eqref{eq:prev} we get $\bm{\mu}_{k,i}(\thetatrue)\xrightarrow{\rm p}1$. Since the network is strongly connected (Assumption~\ref{th:strong}), by iterating the above reasoning we conclude that $\bm{\mu}_{k,i}(\thetatrue)\xrightarrow{\rm p}1$ for any agent, which implies $\bm{m}_i=\sum_{k=1}^Kv_k\log \bm{\mu}_{k,i}(\thetatrue) \xrightarrow{\rm p}0$.
From part $2)$ of Lemma~\ref{th:sumart-true}, this convergence must take place almost surely: $\sum_{k=1}^Kv_k\log \bm{\mu}_{k,i}(\thetatrue) \xrightarrow{\rm a.s.}0$.
Since $v_k>0$ for all $k$, we must have $\bm{\mu}_{k,i}(\thetatrue) \xrightarrow{\rm a.s.}1$ for every agent $k$, which concludes the proof for the case $\rho=0$.

{\em II. Case $\rho>0$.}
In this case $\mathcal{I}_k$ is nonempty for any agent $k$, and \eqref{eq:0} immediately follows from Lemma~\ref{lem:ntxhyp}.
Thus, we focus on proving \eqref{eq:const-lim} and \eqref{eq:residualindistinguish}.

From~\eqref{eq:model-psi} and~\eqref{eq:model-mu} we have, for each agent $k$:
\begin{equation}
	\frac{\bm{\mu}_{k, i} (\mathcal{I}_k)}{\bm{\mu}_{k, i} (\thetatrue)}\!=\! \prod_{\ell = 1}^K \! \left( \frac{\bm{\psi}_{\ell, i}
		(\mathcal{I}_{\ell})}{\bm{\psi}_{\ell, i} (\thetatrue)} \right)^{\!\! a_{\ell k}}\!\!\!\!
	\frac{\bm{\psi}_{k, i} (\mathcal{I}_k)}{1\!-\!\bm{\psi}_{k, i}
		(\thetatrue)}\! \prod_{\ell = 1}^K \!\left( \!\frac{1 \!-\!\bm{\psi}_{\ell, i}
		(\thetatrue)}{\bm{\psi}_{\ell, i} (\mathcal{I}_{\ell})} \right)^{\!\! a_{\ell k}}\!\!\!\!.
	\label{eq:mumu-ind0}
\end{equation}
Using \eqref{eq:model-memorypsi2} and since the likelihood of an indistinguishable hypothesis is equal to the likelihood of the true hypothesis, we get:
\begin{equation}
	\frac{\bm{\psi}_{\ell, i} (\mathcal{I}_{\ell})}{\bm{\psi}_{\ell, i}
		(\thetatrue)} = \frac{L (\bm{\xi}_{\ell, i} | \thetatrue)
		\bm{\mu}_{\ell, i - 1} (\mathcal{I}_{\ell})}{L (\bm{\xi}_{\ell, i} |
		\thetatrue) \bm{\mu}_{\ell, i - 1} (\thetatrue)}
	=
	\frac{\bm{\mu}_{\ell, i - 1} (\mathcal{I}_{\ell})}{ \bm{\mu}_{\ell, i - 1} (\thetatrue)},
\end{equation}
whereas from the definition of distinguishable and indistinguishable sets we have:
\begin{equation}
	\frac{1
		-\bm{\psi}_{\ell, i} (\thetatrue)}{\bm{\psi}_{\ell, i}
		(\mathcal{I}_{\ell})} = \frac{\bm{\psi}_{\ell, i} (\mathcal{I}_{\ell})
		+\bm{\psi}_{\ell, i} (\mathcal{D}_{\ell})}{\bm{\psi}_{\ell, i}
		(\mathcal{I}_{\ell})} 
	=
	1 +
	\frac{\bm{\psi}_{\ell, i}
		(\mathcal{D}_{\ell})}{\bm{\psi}_{\ell, i} (\mathcal{I}_{\ell})}.
\end{equation}
Therefore, if we introduce the vectors:
\begin{align}
	\bm{y}_i &\triangleq {\rm col} \left\{ \log \frac{\bm{\mu}_{k,
			i} (\mathcal{I}_k)}{\bm{\mu}_{k, i} (\thetatrue)} \right\}_{k = 1}^K,
\\
	\bm{x}_i \!&\triangleq\! {\rm col} \! \left\{ \!\log \!\left(\! \left( 1\! +\!
	\frac{\bm{\psi}_{k, i} (\mathcal{D}_k)}{\bm{\psi}_{k, i}
		(\mathcal{I}_k)} \right)^{\!\!-1} \!\!\prod_{\ell = 1}^K \!\! \left( 1 +
	\frac{\bm{\psi}_{\ell, i}
		(\mathcal{D}_{\ell})}{\bm{\psi}_{\ell, i} (\mathcal{I}_{\ell})}
	\right)^{\!\! a_{\ell k}} \right)\! \right\}_{\! k = 1}^{\! K}\!\!\!\!,
	\nonumber\\
	\label{eq:xvecdef}
\end{align}
we can recast \eqref{eq:mumu-ind0} in the vector form $\bm{y}_i = A^{\top} \bm{y}_{i - 1} +\bm{x}_i$.
Unfolding the recursion we get:
\begin{equation}
	\bm{y}_i = (A^i)^{\top} y_0 + \sum_{j = 0}^{i - 1} (A^j)^{\top}
	\bm{x}_{i - j} . 
	\label{eq:rec1}
\end{equation}
Let $V$ be the $K \times K$ matrix whose columns are all equal to the Perron
eigenvector, i.e., $V=v\mathbbm{1}^\top$. This matrix satisfies the following properties for all $i = 0, 1,\ldots$:
\begin{itemize}
	\item The vectors $\bm{x}_i$ in \eqref{eq:xvecdef} are in the null space of $V^{\top}$:\footnote{In fact, using the identity $\sum_k a_{\ell
			k} v_k = v_{\ell}$, we have that  the term $\sum_{k = 1}^K v_k [\bm{x}_i]_k$
		is the logarithm of:
		\begin{equation}
			\prod_{k = 1}^K \left( 1 + \frac{\bm{\psi}_{k, i}
				(\mathcal{D}_k)}{\bm{\psi}_{k, i} (\mathcal{I}_k)} \right)^{- v_k}
			\prod_{\ell = 1}^K \left( 1+  \frac{\bm{\psi}_{\ell, i}
				(\mathcal{D}_{\ell})}{\bm{\psi}_{\ell, i} (\mathcal{I}_{\ell})}
			\right)^{{\sum_k}  a_{\ell k} v_k}  = 1.
	\end{equation}}
	\begin{equation}
		V^{\top} \bm{x}_i = 0.
		\label{eq:Vx0}
	\end{equation}
	\item There exist two constants $\kappa > 0$ and $0 < \beta <1$ such that~\cite[Th. 8.5.1, p. 516]{bib:matrix}:
	\begin{equation}
		\max_{k, \ell = 1, \ldots, K} \big| [A^i - V]_{k\ell} \big| \leq \kappa
		\beta^i . \label{eq:my-favorite-bound}
	\end{equation}
\end{itemize}
In view of~\eqref{eq:Vx0} we can rewrite~\eqref{eq:rec1} as:
\begin{equation}
	\bm{y}_i = (A^i)^{\top} y_0 + \sum_{j = 0}^{i - 1} B_j
	\bm{x}_{i - j},
	\label{eq:unrolled}
\end{equation}
where $B_j \triangleq (A^j - V)^{\top}$. 
Now we prove that:
\begin{equation}
	\Big\| \sum_{j = 0}^{i - 1} B_j
	\bm{x}_{i - j}  \Big\|_{\infty} 
	\xrightarrow{{\rm a.s.}} 0.
	\label{eq:sub-claim}
\end{equation}
Letting $\bm{z}_i\triangleq \kappa \sum_{\ell = 1}^K | [\bm{x}_i]_{\ell} |$, we can write:
\begin{align}
	&\Big\|
	\sum_{j = 0}^{i - 1} B_j \bm{x}_{i -
		j} \Big\|_{\infty} 
		= 
		\max_{k \in [1, K]} \Big| 
		\sum_{j = 0}^{i - 1}
	\sum_{\ell = 1}^K [B_j]_{k\ell}  [\bm{x}_{i - j}]_{\ell} \Big|
	\nonumber\\
	& \leq  \sum_{j = 0}^{i - 1} 
	\max_{k \in [1, K]} \Big(\big| [B_j]_{k\ell} \big| \Big)
	\sum_{\ell = 1}^K 
\big| [\bm{x}_{i - j}]_{\ell} \big|
\leq \sum_{j = 0}^i \beta^j \bm{z}_{i - j}, 
\end{align}
where in the last step we used the bound from~\eqref{eq:my-favorite-bound}.
From Lemma~\ref{th:mart00} it is readily seen that $[\bm{x}_i]_\ell \xrightarrow{{\rm a.s.}}
0$ for any $\ell$, and, hence, $\bm{z}_i \xrightarrow{{\rm a.s.}} 0$. This implies that, almost surely, for any $\varepsilon > 0$ there exists a (random) value
$\bm{N}_{\varepsilon}$ such that for each $i > \bm{N}_{\varepsilon}$ we have
$\bm{z}_i < \varepsilon(1 - \beta)$. Therefore, we can write:
\begin{align}
	\sum_{j = 0}^i \beta^j \bm{z}_{i - j} &= \sum_{j = 0}^{i - \bm{N}_{\varepsilon} - 1} \beta^j \bm{z}_{i - j}
	+ \sum_{j = i - \bm{N}_{\varepsilon}}^i \beta^j \bm{z}_{i - j} \nonumber\\ &< \varepsilon + \sum_{j = i - \bm{N}_{\varepsilon}}^i \beta^j \bm{z}_{i - j}\qquad {\rm a.s.,}
	\label{eq:decomposition}
\end{align}
where in the inequality we also used the fact that $\sum_{j=0}^\infty \beta^j=\frac{1}{1-\beta}$.
Moreover, since $\beta^i$ is decreasing, we get:
\begin{equation}
	\sum_{j = i - \bm{N}_{\varepsilon}}^i \beta^j \bm{z}_{i - j} <
	\beta^i \frac{1}{\beta^{\bm{N}_{\varepsilon}}} \sum_{j = 1}^{\bm{N}_{\varepsilon}}
	\bm{z}_j \triangleq \beta^i \bm{w}_\varepsilon,
\end{equation}
where $\bm{w}_{\varepsilon}$ is almost-surely finite in view of Assumptions~\ref{th:finiteKL} and~\ref{th:positivemu0}.
Thus, from~\eqref{eq:decomposition} we obtain:
\begin{equation}
	\limsup_{i\rightarrow\infty} \sum_{j=0}^i \beta^j \bm{z}_{i-j} \leq \varepsilon\qquad {\rm a.s.,}
\end{equation}
which proves~\eqref{eq:sub-claim} in view of the arbitrariness of $\varepsilon$. Since from~\eqref{eq:my-favorite-bound}:
\begin{equation}
	[(A^i)^{\top} y_0]_k \xrightarrow{i \rightarrow \infty} \left[V^{\top} y_0\right]_k = \log \frac{1}{\rho},
\end{equation}
where we have applied the definition of the network confusion ratio $\rho$ --- see \eqref{eq:confunet}.
Then, Eqs.~\eqref{eq:unrolled} and~\eqref{eq:sub-claim} imply:
\begin{equation}
	\frac{\bm{\mu}_{k,i}(\mathcal{I}_k)}{\bm{\mu}_{k,i}(\thetatrue)} \xrightarrow{{\rm a.s.}}  \frac{1}{\rho}.
	\label{eq:limit}
\end{equation}
On the other hand, from~\eqref{eq:0} we have:
\begin{equation}
	\bm{\mu}_{k, i} (\thetatrue) +\bm{\mu}_{k, i} (\mathcal{I}_k)
	\xrightarrow{\rm a.s.} 1,
	\label{eq:lim1}
\end{equation}
and since we can write:
\begin{equation}
	\bm{\mu}_{k, i} (\thetatrue) +\bm{\mu}_{k, i} (\mathcal{I}_k)
	=
	\bm{\mu}_{k, i} (\thetatrue) \left(1+
	\frac{\bm{\mu}_{k,i}(\mathcal{I}_k)}{\bm{\mu}_{k,i}(\thetatrue)}
	\right),
\end{equation}
the convergences in~\eqref{eq:limit} and~\eqref{eq:lim1} imply~\eqref{eq:const-lim} and \eqref{eq:residualindistinguish}.
Finally, for each pair of indistinguishable hypotheses $\theta, \tau \in \mathcal{I}_k$:
\begin{equation} 
	\frac{\bm{\mu}_{k, i} (\tau)}{\bm{\mu}_{k, i} (\theta)}
	= \frac{\mu_{k, 0} (\tau)}{\mu_{k, 0} (\theta)},
	\label{eq:indishable-propto}
\end{equation}
and \eqref{eq:1-c} follows from \eqref{eq:subsetS}. 



\end{document}